\documentclass[twocolumn]{aastex631}

\usepackage{lineno}
\usepackage{natbib}

\newcommand{\axaf}{\mbox{\em Chandra\/}}
\received{ }
\revised{  }
\accepted{  }
\submitjournal{ApJ}

\shorttitle{X--ray astrometry with mas precision at $z>1$ using strong lensing }
\shortauthors{Spingola, Schwartz \& Barnacka}
\graphicspath{{./}{figures/}}

\begin{document}


\title{Milliarcsecond X-ray astrometry to resolve inner regions of AGN at $z>1$ using gravitational lensing}

\correspondingauthor{Cristiana Spingola}
\email{spingola@ira.inaf.it}

\author[0000-0002-2231-6861]{Cristiana Spingola}
\affiliation{INAF $-$ Istituto di Radioastronomia, Via Gobetti 101, I$-$40129, Bologna, Italy}

\author[0000-0001-8252-4753]{Daniel Schwartz}
\affiliation{Smithsonian Astrophysical Observatory, Cambridge, MA 02138, USA}

\author[0000-0001-5655-4158]{Anna Barnacka}
\affiliation{Smithsonian Astrophysical Observatory, Cambridge, MA 02138, USA}
\affiliation{Astronomical Observatory, Jagiellonian University, ul. Orla 171, 30-244 Cracow, Poland}



\begin{abstract}

We report the localization of the X-ray emission from two strongly lensed AGN, CLASS B0712+472 ($z=1.34$) and CLASS B1608+656 ($z=1.394$).  We obtain milliarcsecond X-ray astrometry by developing a novel method that combines parametric lens modelling with a Bayesian analysis. We spatially locate the X-ray sources in CLASS B0712+472 and CLASS B1608+656 within 11 mas and 9  mas from the radio source, respectively.  For CLASS B0712+472, we find that the X-ray emission is co-spatial with the radio and optical emission.  While, in CLASS B1608+656, the X-ray emission is co-spatial with radio, but displaced with respect to the optical emission at 1$\sigma$ level, which positions this source as an offset AGN candidate.  This high astrometric precision improves on the limitations of existing X-ray instruments by two orders of magnitude.  The demonstrated method opens a path to search for offset and binary AGN at $z>1$, and to directly test supermassive black hole formation models in a redshift range that has been mostly underconstrained to date.

\end{abstract}



\section{Introduction} \label{sec:intro}

Active galactic nuclei (AGN) are supermassive black holes (SMBHs) actively accreting the surrounding gas and stars \citep[e.g.,][]{Padovani2017}. They are among the most energetic sources in the Universe, hence they have a strong impact in the shaping and evolution of their host galaxies  \citep[e.g.,][]{Fabian2012}. Understanding the properties of AGN at all redshifts is crucial to assess how they form and what is their role in the evolution of galaxies across the cosmic time.

Binary SMBH systems are a natural consequence of the current structure formation scenario, where the galaxy building is driven by a hierarchical process of merging \citep[e.g.,][]{Hopkins2008, Somerville2015}. These AGN pairs can be identified as two distinct flat-spectrum radio sources, two X-ray sources, through multiple peaks of the narrow emission lines or via drops in the broad-band spectral energy distribution of an AGN due to a gap in the accretion disk caused by the secondary SMBH \citep[e.g.,][]{Burke-Spolaor2011, Koss2012, Fu2012, Gultekin2012, Yan2015}.  We follow the nomenclature of \citet{Burke-Spolaor2014} and define \textsl{dual AGN} those SMBH pairs separated by $<10$ kpc, while \textsl{binary AGN} are closer pairs of SMBHs separated by $<100$ pc,  which approximately corresponds to the Bondi radius of a SMBH of $5\times10^8$ M$_{\odot}$.

In these SMBH pairs it is possible that only one is active. In this case, the binary system may be observed as an offset-AGN, as the radio/X-ray radiation is offset with respect to the peak of the optical emission of the host galaxy \citep{Orosz2013, Lena2014, Kim2016, Barrows2016, Barrows2018, Skipper2018}. Nevertheless, offset-AGN are not necessarily binary SMBHs, but they could also be a single recoiling SMBH, namely a SMBH that has been displaced from the center during the merging process \citep{Madau2004, Volonteri2008, Lena2014}. Both binary and offset-AGN can provide important clues on the fraction of galaxy mergers across the cosmic time \citep[e.g.,][]{Comerford2012, Silva2021}.

It is predicted that the merging of SMBHs would produce loud gravitational waves at the frequency range nHz$-\mu$Hz, which is the observing range of the Laser Interferometer Space Antenna (LISA) and the Pulsar Timing Array (PTA, e.g., \citealt{Enoki2004, Burke-Spolaor2019, DeRosa2019}). Even though, currently, the observed number of offsets and binary SMBHs seems to fit the theoretical expectations from the $\Lambda$CDM model, such sample is limited mostly to low-$z$ SMBHs \citep[e.g.,][]{Rosas-guevara2019, Bartlett2021}.  However, galaxy mergers are more common at early times \citep[e.g.,][]{Conselice2003, Hopkins2006}, so we expect to observe more frequently offset and binary SMBHs at high redshifts. To fully test the hierarchical cosmological model and characterize the primary sources of LISA and PTA is, therefore, necessary to assess the occurrence of offset/binary SMBHs at large distances.

The main issues for the detection and identification of offset and binary AGN systems are the high sensitivity and angular resolution required to spatially resolve them on pc-scales. Hence, at high redshift spatially resolving binary AGN is even more challenging, requiring milliarcsecond angular resolution and long exposure times. Given the paucity of radio-loud AGN that can be observed with very long baseline interferometry (VLBI), to obtain a representative sample the ideal observing band is at X-rays. However, X-ray telescopes have very limited angular resolution (\axaf\ resolution is $\sim$0.5 arcsec), which makes the identification of such systems at large redshifts impossible. 

Gravitational lensing can provide the necessary amplification and magnification for studying the pc-scales of distant background sources \citep{Barnacka2017, Barnacka2018}. Gravitational lensing consists of the deflection of light from a distant background source by a foreground massive object (called lens; \citealt{Congdon2018}). As a consequence of galaxy-galaxy lensing, multiple magnified (and distorted) images of the same high-redshift source may be observed to be separated on scales of $>$1 arcsec, because the typical mass of a lensing galaxy is $M \sim 10^{11}$ M$_{\odot}$ \citep{Koopmans2006, Auger2009}. These scales are resolvable, for instance, at X-rays with the \axaf\ telescope, at optical wavelengths with the Hubble Space Telescope (HST) and at the radio wavelengths with VLBI arrays. Once corrected for the distortion due to lensing, it becomes then possible to recover the multi-wavelength emission of distant sources on sub-galactic scales  \citep{Deane2013, Spilker2015, Barnacka2015, Barnacka2016, Massardi2018, Dye2018, Spingola2019, Spingola2020_gas, Rybak2020, Berta2021} and search for offset/binary AGN systems \citep{Spingola2019, Spingola&Barnacka2020, Schwartz2021}.

Here, we present an innovative method to push the limitations of the X-ray telescopes and constrain at high precision the location of the X-ray emission to search for offset and binary AGN systems. 
As a pilot study \citet{Spingola&Barnacka2020} selected two gravitationally lensed sources in a so-called ``caustic configuration" (i.e. quadruply imaged sources, \citealt{Barnacka2017}).  
Gravitational lenses with elliptical mass distribution deflect the light of background sources that when ray-traced back to the source plane form a shape of a characteristic caustics (i.e. points of infinite magnification). 
The diameter of such a caustic in the source plane depends on the lens ellipticity and scales with the Einstein radius. Sources positioned in the proximity of the inner side of such caustic experience significant magnification of lensed images and amplification of the image positions in respect to the location of the source, \citep{Barnacka2017, Barnacka2018}.
Since these specific lensed sources are at high magnification, they can provide a promising sample to search for offset and binary AGN over a wide range in spatial separations, including the crucial sub-kpc scales (where the SMBHs pairs are gravitationally bound), which are challenging to reach at high redshifts. We use these caustics as non-linear spatial amplifiers, which allows us to connect to the International Celestial Reference Frame established by radio observations, thus overcoming the technological limitations of existing instruments.

The paper is structured as follows. In  Sec.~\ref{sec:observations} we describe the two targets and the X-ray observations. In Sec.~\ref{sec:method} we provide the details of our methodology to recover the X-ray source position and its uncertainty. We then present our results (Sec.~\ref{sec:results}), a discussion of the results (Sec.~\ref{sec:discussion}), and conclusions (Sec.~\ref{sec:conclusions}). We adopt the \citet{Planck2016} cosmological values, namely $H_0=67.8$ km s$^{-1}$ Mpc$^{-1}$, $\Omega_M = 0.308$ and $\Omega_{\Lambda}= 0.692$. This set of cosmological parameters gives a scale of 8.626 pc mas$^{-1}$ at $z=1.34$ and 8.653 pc mas$^{-1}$ at $z=1.394$.

\section{Chandra Observations} \label{sec:observations}

\begin{figure*}
    \centering

    \includegraphics[width=\textwidth]{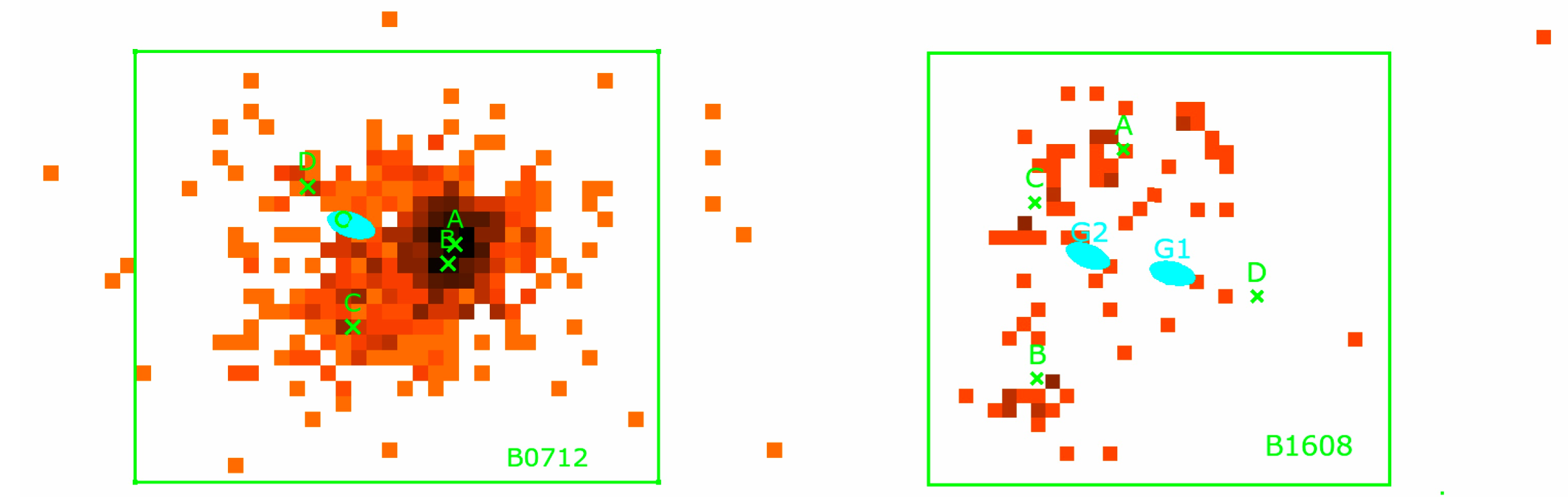}
    \caption{ X-ray data in the 0.5 -- 7 keV band, binned into 1/4 pixels, 0\farcs123 for both sources. The square shows the data region used for the maximum likelihood calculation. CLASS B0712+472 (left panel) uses an 8.5$\times$ 7 pixel region (4\farcs18$\times$3\farcs44) with a peak count of 26 per bin in black, and orange representing 1 count. We expect 1.9 background counts in this region.  CLASS B1608+656 (right panel) uses an 8$\times$7.5 (3\farcs93$\times$3\farcs69) pixel region with a peak count of 3 per bin in dark brown, and red representing 1 count per bin. We expect 0.62 background counts in this region. The \emph{x's} indicate the positions of the VLBI images with the notation in \citet{Spingola&Barnacka2020}, while the cyan ellipses indicate their positions for the lensing galaxies. However, note that there is a systematic offset of order  0\farcs5 due to the \axaf\ absolute celestial location uncertainty. More details can be found here \href{https://cxc.harvard.edu/cal/ASPECT/celmon/}{https://cxc.harvard.edu/cal/ASPECT/celmon/}.
    }
    \label{fig:databins}
\end{figure*}

\begin{figure*}
    \centering
        \includegraphics[width=\textwidth]{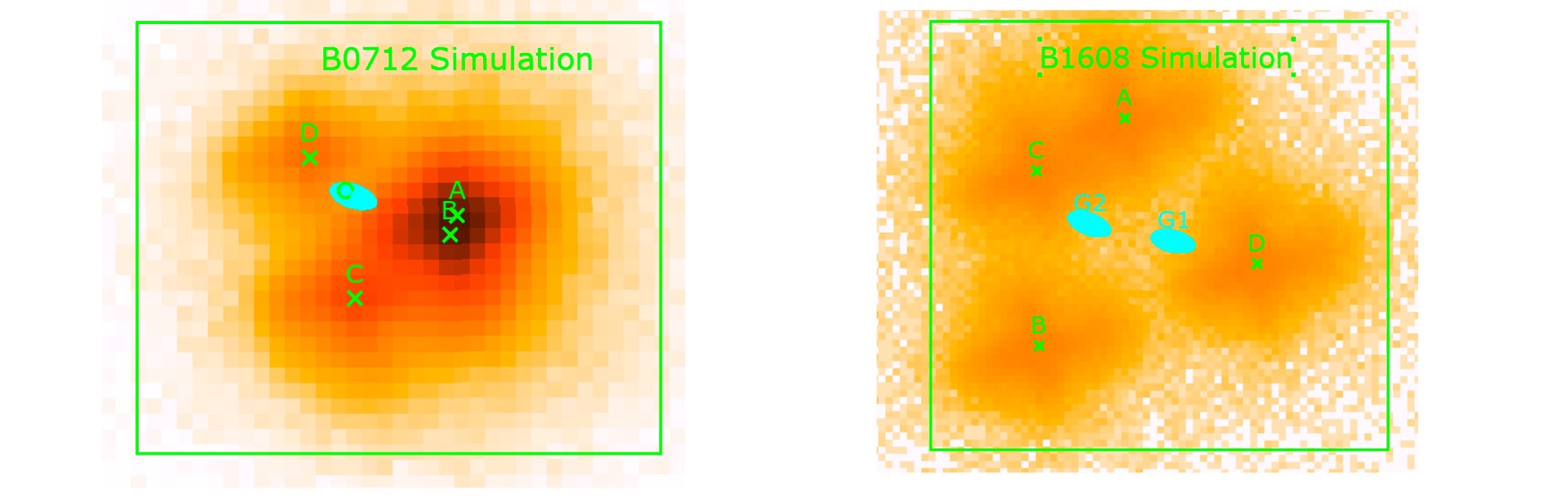}
    \caption{Simulation of the four point source images, in 1/4 pixels covering the same regions and the 0.5 -- 7 keV band as in Fig. \ref{fig:databins}. The \textit{x's} and cyan ellipses are the same as in Figure 1. \textsl{Left:} 696,384 simulated photons for the four point lensed images of CLASS B0712+472. Multiply by 0.0012 to equal the total observed counts. The maximum pixel has 18941 counts, while the faintest color pixels have 10 -- 20 counts. The maximum likelihood position fit to the data gives 1.69, 0.73, 0.34, and 1.01 times the normalized predicted lens model amplitudes to the fluxes of images A, B, C, D, respectively. \textsl{Right:} 161,044 simulated photons for the four point lensed images of CLASS B1608+656. Multiply by 0.00049 to equal the total observed counts. The maximum pixel is in image A and has 313 photons. The faintest color has 4 photons, white has 0 -- 3. The maximum likelihood fit gives fluxes that are 1.75, 0.51, 1.22 and 1.25 times the normalized predicted amplitudes from the lens mass model.
    }
    \label{fig:psf_simulation}
\end{figure*}

We apply our novel method to two flat-spectrum radio-loud gravitationally lensed AGN: CLASS~B0712+472 and CLASS~B1608+656. These systems are both quadruply imaged. A detailed description of the radio and optical properties and lens modelling analysis of the targets can be found in \citet{Spingola&Barnacka2020}. Here, we summarize the key characteristics and focus on X-ray emission.

\subsection{CLASS~B0712+472} 
The HST emission of the lensed images of CLASS~B0712+472 is detected at high significance, especially the most magnified images  A and B \citep{Hsueh2017,Spingola&Barnacka2020}. The quasar is at $z=1.34$, and the lensing galaxy at $z=0.406$ \citep{Fassnacht1998}. Adaptive optics observations at NIR wavelengths showed that there is also a faint diffuse emission into arcs, which is likely due to the dust and stellar emission of the AGN host galaxy \citep{Hsueh2017}. \axaf\ observations of CLASS B0712+472 were carried out on 2003 December 17 (ObsID 4199) for 97.7 ks live time \citep{Fassnacht2008}, for the purpose of studying a foreground group of galaxies at $z=0.29$.
See also \citet{Momcheva2015, Wilson2017}, for other data on these groups, concurring that the group at $z=0.29$ had a minimal effect on the lensed images as reported by \citet{Fassnacht2002}.  \citet{Fassnacht2008} reported X-ray detection of three images out of the four, with images A and B blended together due to their small angular separation of about 0\farcs2. 

We have used CIAO$-$4.14 \citep{Fruscione2006} with CALDB 4.9.6 to reanalyze the REPRO 4 \axaf\ X-ray data from the ObsID 4199.  We find 856 photons between 0.5 and 7 keV in the rectangular region shown in the left of Figure~\ref{fig:databins}. Fitting to a power law with foreground absorption of n$_H$=7.26$\times$10$^{-20}$ H-atoms cm$^{-2}$ in our Galaxy \citep{Dickey1990} gives an energy index\footnote{We use the convention that the flux density $f_{\nu}\propto\nu^{-\alpha}$.} $\alpha=0.524 \pm 0.060$ giving an unabsorbed energy flux of (7.28$\pm$0.32)$\times$10$^{-14}$ erg s$^{-1}$ cm$^{-2}$. 

Separately fitting 595 photons in an 0\farcs7 radius circle centered on the blended A/B location, gives a fit with $\alpha=0.54\pm 0.07$. Fitting the remaining photons gives $\alpha=0.47\pm0.11$, so we conclude that all four images are consistent with the flat spectral index $\alpha=0.52$.  Absorption can also be present in the lensing galaxy, especially if late-type \citep{Dai2009}.
Allowing for intrinsic absorption at the quasar redshift $z=1.34$ gives a more typical quasar index of $\alpha=0.64\pm 0.10$, with an intrinsic column density fit to (4.1$\pm$2.2)$\times$10$^{22}$ H-atoms cm$^{-2}$. \citet{Wang2016} reported a total flux in the 0.3 to 8 keV band of 9.8 $\times$10$^{-14}$ erg s$^{-1}$ cm$^{-2}$ based on assuming that the spectral index was $\alpha=0.7$. We derive 6.9 $\times$10$^{-14}$ erg s$^{-1}$ cm$^{-2}$ for the same energy band if we fix the spectral index to be $\alpha=0.7$. The different contamination models on the ACIS filter in 2003 can possibly account for these differences.

The analysis of the optical and radio observations in the source plane has shown that the two emissions are co-spatial, with a relative offset of $17\pm42$ pc \citep{Spingola&Barnacka2020}. This astrometric precision at $z=1.34$ could be achieved thanks to the sensitivity of the observations and a magnification factor $\mu > 10$.
 
\subsection{CLASS~B1608+656}

The source in the lensing system CLASS~B1608+656 consists of a post-starburst galaxy located at $z_s = 1.394$, which is magnified by two lensing galaxies at $z_l = 0.630$ \citep{Fassnacht1996}.  CLASS~B1608+656 shows strong flux density variability at both radio and optical wavelengths \citep[e.g.,][]{Fassnacht1999}.
\citet{Dai2005} used \axaf\ to observe CLASS~B1608+656 for 29.7 ks on 21 September 2003 (ObsID 3461). They detected three of the lensed images, with the fourth being too faint. The authors did not find an X-ray emission that could be associated with a group (or a cluster) associated with the two main lensing galaxies. \citeauthor{Dai2005} also modelled the 0.4 to 8 keV spectrum of CLASS~B1608+656, and used $\chi^2$ to fit a power-law photon index of $\Gamma = 1.4\pm0.3$, a Galactic absorption of $N_H < 6.4\times10^{20}$ cm$^{-2}$ and upper limits on absorption column at the AGN redshift of $N_H < 2 \times 10^{21}$ cm$^{-2}$. 
The observed X-ray flux ratios were broadly consistent with those at the radio wavelengths. 

The post-starburst spectrum of CLASS~B1608+656 could be linked to a merger event that is responsible for the radio-optical offset found by the VLBI--HST source plane analysis of \citet{Spingola&Barnacka2020}. The authors found a radio-optical offset of $214\pm137$ pc, which makes this source a promising offset-AGN candidate at high redshift to be investigated at X-ray wavelengths.

We reanalyzed the \axaf\ observation of CLASS B1608+656 (ObsID 3461, PI: Kochanek), omitting a short 8.1 ks (ObsID 429) observation from January 2000.  We used CIAO-4.12 \citep{Fruscione2006} and the REPRO 4 data, yielding 79 photons in the 0.5--7 keV band for CLASS B1608+656. Our power law fit, with galactic absorption fixed at 2.68 $\times 10^{20}$ cm$^{-2}$ \citep{Dickey1990} gave $\alpha=0.43\pm0.21$, and an energy flux in the 0.5 to 7 keV band of (2.99$\pm$0.44)$\times$10$^{-14}$ erg s$^{-1}$ cm$^{-2}$.  The \axaf\ image for this system is shown in the right panel of Fig.~\ref{fig:databins}.

\section{Methodology} \label{sec:method}

In this section we describe our novel method to locate at high astrometric precision the X-ray emission in the source plane. The method can be divided in two main steps: simulation of the X-ray source positions (Sec.~\ref{sec:lensmodel}) and  maximum likelihood plus correlation analysis in the lens plane to identify the best set of lensed images associated with a specific source location (Sec.~\ref{sec:likelihood}).

\subsection{Lens models and simulated source positions}\label{sec:lensmodel}

We performed lens mass modelling to simulate the range of the X-ray source position. The model corrects for the lensing distortion by applying a parametrized profile for the mass density distribution of the deflector. For the investigated systems, the lens mass models were inferred from VLBI and HST observations, thus, providing high fidelity source reconstruction. The lens mass model parameters for CLASS B0712+472 and CLASS B1608+656 have uncertainties less than 10 \% \citep{Koopmans2003,Hsueh2017, Spingola&Barnacka2020}. All the lens mass model parameters and their uncertainties are reported in Table 5 of \citet{Spingola&Barnacka2020}. The mass density distribution of the lens of CLASS B0712+472 is parametrized by two mass components (an elliptical power-law plus an exponential disk), because the lensing galaxy is a late-type galaxy \citep{Hsueh2017}. The lens mass model of CLASS B1608+656 consists of two elliptical power-laws, as there are two main lensing galaxies, with no evidence of disk structure \citep{Koopmans2003, Suyu2010}.

We use the VLBI source position (Table 6 of \citealt{Spingola&Barnacka2020}) as a reference point to generate equally spaced source positions in the perpendicular and parallel directions with respect to the caustics.  The actual spacing will depend on the statistical significance we can achieve with the given number of X-ray photons.  For CLASS B0712+472 and CLASS B1608+656 we take lines of possible source positions spaced 1 mas and 5 mas, respectively, to determine the initial locations of the X-ray sources (Figures \ref{fig:b0712_simulated_sources} and \ref{fig:b1608_simulated_sources}).

\subsection{Maximum likelihood plus correlation analysis}\label{sec:likelihood}

In this section we describe our general method (briefly introduced in \citealt{Schwartz2021}) that can be applied to any multiply imaged lensed sources, but we add with some notes specific to the sources CLASS B0712+472 and CLASS B1608+656 reported on in this article.

\begin{figure*}
    \centering
    \includegraphics[width=\textwidth]{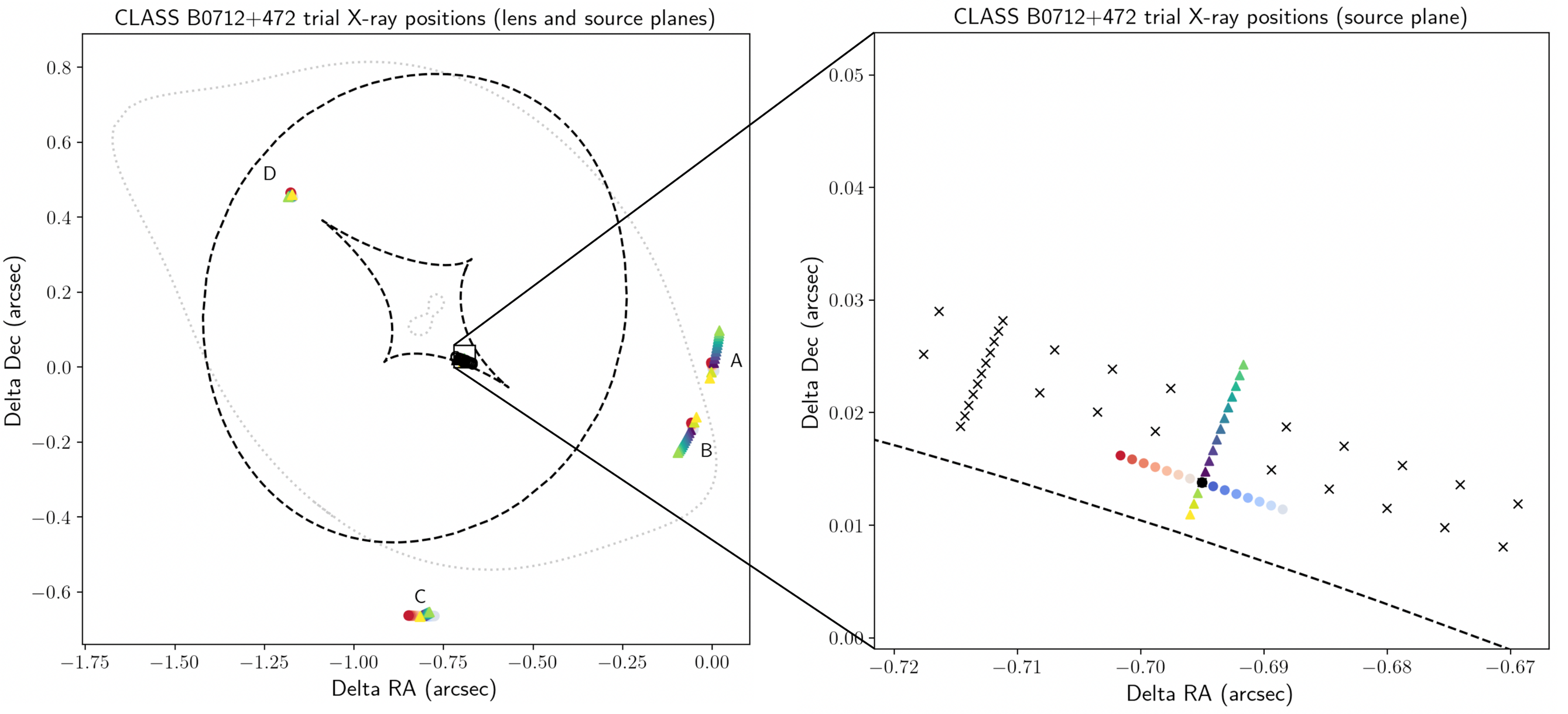}
    \caption{Set of four lensed images (left) associated with the trial sources separated by 1 mas (right, same color and symbol of the corresponding lensed images) for the lensing system CLASS B0712+472. The lens critical curve is shown by the grey dotted line, while the source plane caustics are indicated by the dashed black lines. The triangles indicate trial sources perpendicular to the inner caustic line, while the filled circles indicate trial sources parallel to the caustic. The crosses indicate additional points chosen to define the contours shown in Fig. \ref{fig:source_reconstruction_b0712}. The VLBI source position is indicated by a black filled circle. All the positions are relative to the VLBI position of image A, which is at 07$^h$16$^m$3.576$^s$, $+$47$^{\circ}$08'50.154''.}  
    \label{fig:b0712_simulated_sources}
\end{figure*}

\begin{figure*}
    \centering
    \includegraphics[width=\textwidth]{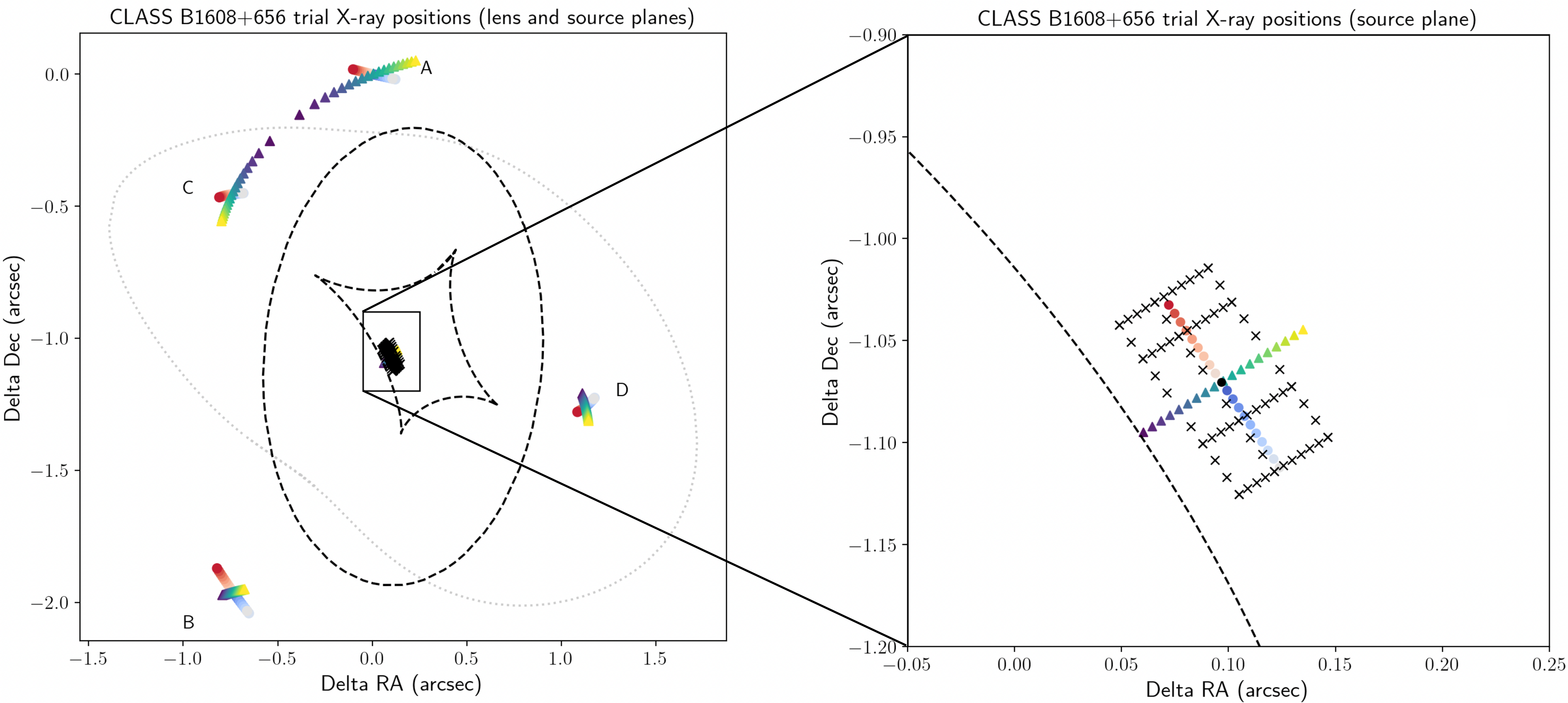}
    \caption{Set of four lensed images (left) associated with the trial sources separated by 5 mas (right, same color and symbol of the corresponding lensed images) for the lensing system CLASS B1608+656. The lens critical curve is shown by the grey dotted line, while the source plane caustics are indicated by the dashed black lines. The triangles indicate trial sources perpedicular to the inner caustic line, while the filled circles indicate trial sources parallel to the caustic. The VLBI source position is indicated by a black filled circle. The crosses indicate additional points chosen to define the contours shown in Fig. \ref{fig:source_reconstruction_b1608}. All the positions are relative to the VLBI position of image A, which is at 16$^h$09$^m$13.956$^s$, $+$65$^{\circ}$32'28.995".}    \label{fig:b1608_simulated_sources}
\end{figure*}

{\bf Step 1:} We simulate the \axaf\ point source response numerically, for each observation of each lensed image (Fig. \ref{fig:psf_simulation}). We run 1000 simulations incorporating the actual observation duration, aspect solution, dither, the estimated source flux, and charge coupled device (CCD) event pileup from the observation, and merge the results into a single fits file. We use SAOTrace \citep{Jerius2004} for a high fidelity mirror simulation, and Marx \citep{Davis2012} for a high fidelity simulation of the ACIS image, including the energy dependent sub-pixel event repositioning. The raytrace takes the celestial position of each image from the radio position given in Table 3 or Table 4 of \citet{Spingola&Barnacka2020}, respectively, for the four images (A, B, C, D) of CLASS B0712+472 or CLASS B1608+656.

\indent {\bf Step 2:} We use the lensing mass model (described in Section~\ref{sec:lensmodel}) to perform a forward-ray tracing using \textsc{gravlens} \citep{Keeton2001a, Keeton2001b} of the trial source positions, predicting the separations and magnifications of sets of four lensed images associated with each test source, while keeping the lens mass model parameters fixed to those from \citep{Spingola&Barnacka2020}. These are shown in Figures \ref{fig:b0712_simulated_sources} and \ref{fig:b1608_simulated_sources}.
The uncertainties on the lens mass model parameters lead to positional differences of less than 1 mas \citep{Spingola&Barnacka2020}, therefore negligible for the fitting process of the X-ray lensed images.
In a general case we might use a finer grid spacing perpendicular to the caustic where the magnification gradient is very steep, while a coarser grid might be taken parallel to the caustic. We add additional lines of positions parallel and perpendicular to the caustic locally, to better define the confidence contours of the allowed locations. In a more general case one might fill in a two dimensional grid either at regular points, or at a sparse or randomly sampled array, depending on how quickly the best fit statistic is changing. 

\indent {\bf Step 3:} We construct a series of models for the predicted X-ray counts by placing the simulated point source images from Step 1 at the predicted lens plane separations and with the predicted relative intensities, for each simulated source position in Step 2. 

\indent {\bf Step 4:} We bin the observed X-ray data into an array of j$\times$k square bins, each with $n_i$ counts. For these two systems, we used 1/4 the CCD pixel size, 0\farcs123. This grid remains fixed with n$_i$ observed counts in each bin. We extend the array about 1\arcsec\  beyond the apparent positions of the X-ray images in the observed data, since our model is sensitive to pixels which are expected to contain only background counts. Figure~\ref{fig:databins} shows the regions we used for these two AGN.

{\bf Step 5:} We normalize each quartet of images and we raster our model in two dimensions using steps of 1/8 of the bin size used in Step 4. Then, for each simulated quartet, we increment the position in the lens plane by this step size, sort into the bins used for the observed data array and add the expected background to predict the expected counts, $\lambda_i$, for each of the j$\times$k  data bins.

\indent {\bf Step 6:} At each of the trial source plane positions, we estimate the maximum likelihood of the simulated data to the observed data. We consider the maximum likelihood for observing the counts in each bin based on Poisson statistics. From the likelihood probability

\begin{equation}
P(\lambda, n)=\prod_i(exp(-\lambda_i) \lambda_i^{n_i}/n_i!)
\end{equation}
we compute
\begin{equation}
C=-2\ln{P} =-2 \sum_{i=1}^{j\times k}\ln{(\lambda_i^{n_i} \exp(-\lambda_i)/n_i!)}.
\end{equation}
We allow the amplitudes of each of the four images to vary by factors consistent with the unknown possibility of ratio anomalies due to microlensing or source variability.  We renormalize the total simulated counts to equal the observed number of counts, M=$\sum_{i=1}^{j\times k}\lambda_i$.   \\
\indent {\bf Step 7:} We linearly interpolate between raster coordinates to find the values of the image A focal plane coordinates which minimize the likelihood
\begin{equation}
   C =2 \left( M-\sum_{i=1}^{j\times k}(n_i\ln{\lambda_i}) \right),
\end{equation}
for each trial source position. As a function of the unknown source position along each one dimensional line, we define confidence regions using the theorem by \cite{Wilks1938} that proves the quantity $\Delta$C (i.e., the likelihood ratio) is distributed as $\chi^2$ almost independent of the underlying density distribution for the observed counts \cite{Cash1979}. The results for location along each line are shown in the top panels of Figures~\ref{fig:source_reconstruction_b0712} and~\ref{fig:source_reconstruction_b1608}, with the confidence limits for one interesting parameter. 

{\bf Step 8:} We take the likelihood at each point of all the one-dimensional lines of trial source positions, and interpolate and extrapolate to create a piece-wise continuous function in two dimensions. For the two interesting parameters of $\Delta$RA and $\Delta$Dec relative to image A, we create the two dimensional contours of statistical confidence about the minimum. Results are shown in the bottom panels of Figures~\ref{fig:source_reconstruction_b0712} and~\ref{fig:source_reconstruction_b1608}. \\

As an additional test, we measured the position of each lensed image and then backward ray-traced the X-ray emission to the source plane. We found that the angular separation of the lensed images for both systems is very different from what is measured using the high angular resolution observations in hand (i.e., optical HST and radio VLBI data). As a result, the optical/radio-derived lens mass models could not reproduce the X-ray images, which is peculiar considering the poor astrometric precision of Chandra data. Typically, such astrometric anomalies for galaxy-galaxy lensing are small (mas-scales) and may be detected only using high angular resolution data (i.e., VLBI), which have the astrometric sensitivity to detect small offsets due to non-smooth mass distributions \citep{Chen2007, Spingola2018, Hartley2019}. Instead, the X-ray lensed images are expected to be well in agreement with a smooth mass density model. As the optical, radio and X-ray emissions are lensed by the same massive object, we concluded that the poor astrometric precision of Chandra observations does not allow us to infer robustly the position of the lensed images, which is to be expected as the merging images are extremely close (hundreds of mas) and image D is faint and difficult to locate in both systems.

\section{Results}\label{sec:results}

\begin{figure*}
    \centering
    \includegraphics[width=0.7\textwidth]{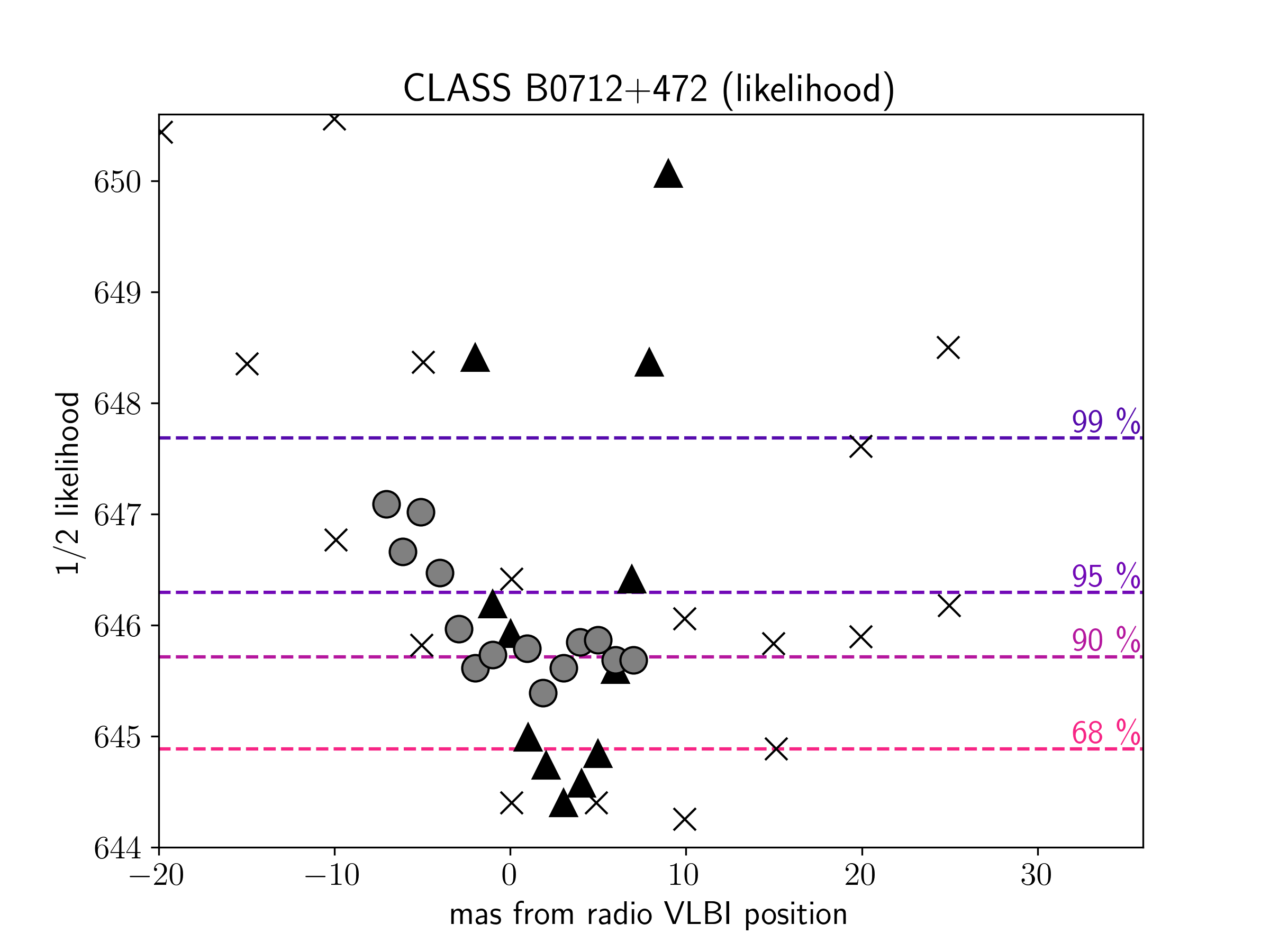} 
    \includegraphics[width=0.7\textwidth]{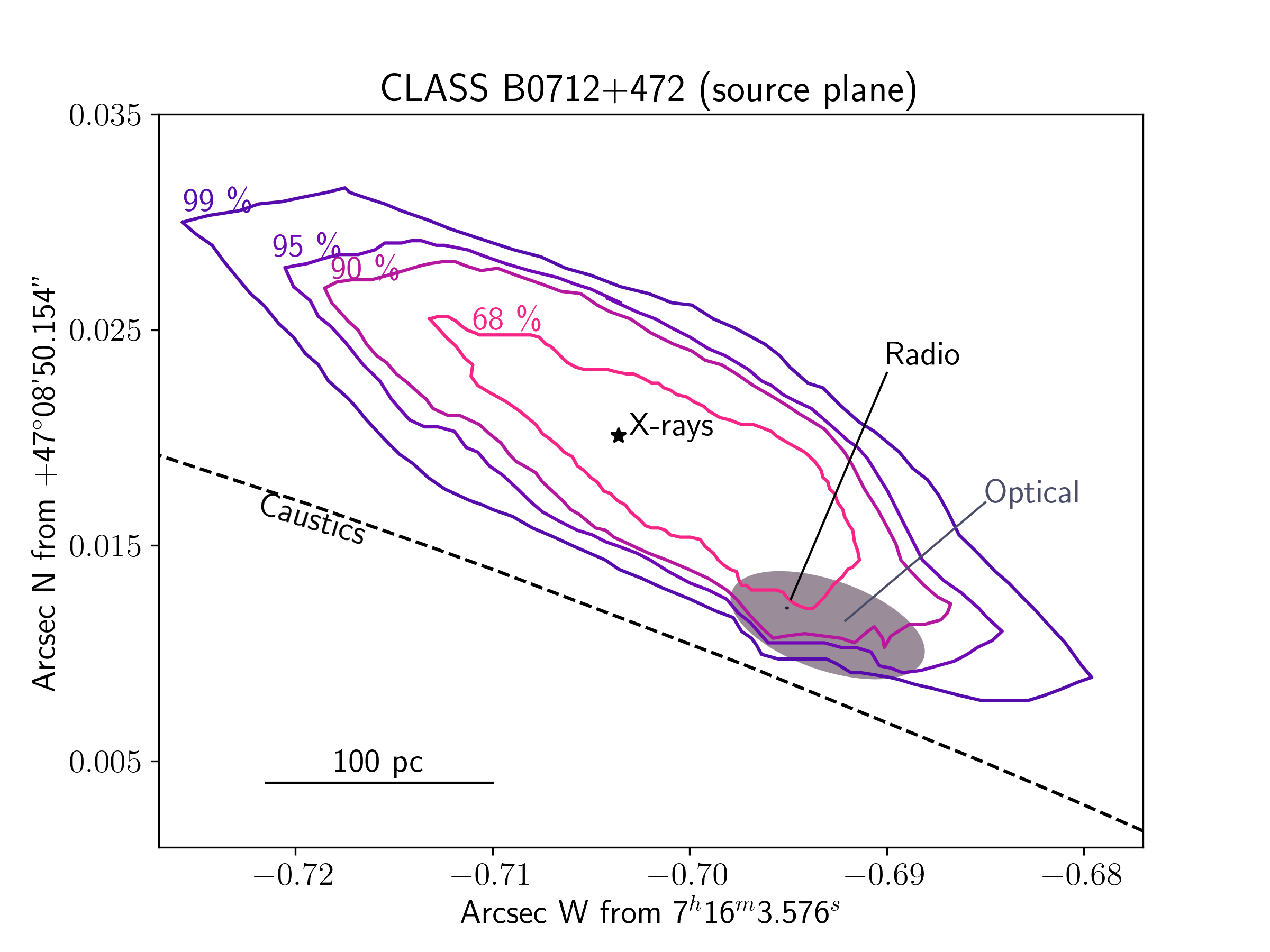}
    \caption{\textsl{Top}: Likelihood values for different trial source positions of CLASS B0712+472. In all cases, black filled triangles are along lines perpendicular to the caustics and grey circles are along lines parallel to the caustics, which correspond to the points shown in Figure~\ref{fig:b0712_simulated_sources}. The black crosses are additional lines of points chosen to define the contours. \textsl{Bottom}: Source plane reconstruction of the X-ray (black star), radio VLBI (black ellipse) and the peak of optical HST (gray ellipse) emissions in CLASS B0712+472. The size of the ellipses indicate the 1$\sigma$ uncertainty. The radio VLBI and HST source locations are from \citet{Spingola&Barnacka2020}. The contours enclose the possible X-ray source position with 99\%, 95\%, 90\% and 68\% confidence, from outer to inner, respectively. The black dashed line shows a portion of the caustic. We highlight that the elongation of the contours is caused by the almost constant magnification factor in the direction parallel to the caustic line. }
    \label{fig:source_reconstruction_b0712}
\end{figure*}

\begin{figure*}
    \centering
    \includegraphics[width=0.7\textwidth]{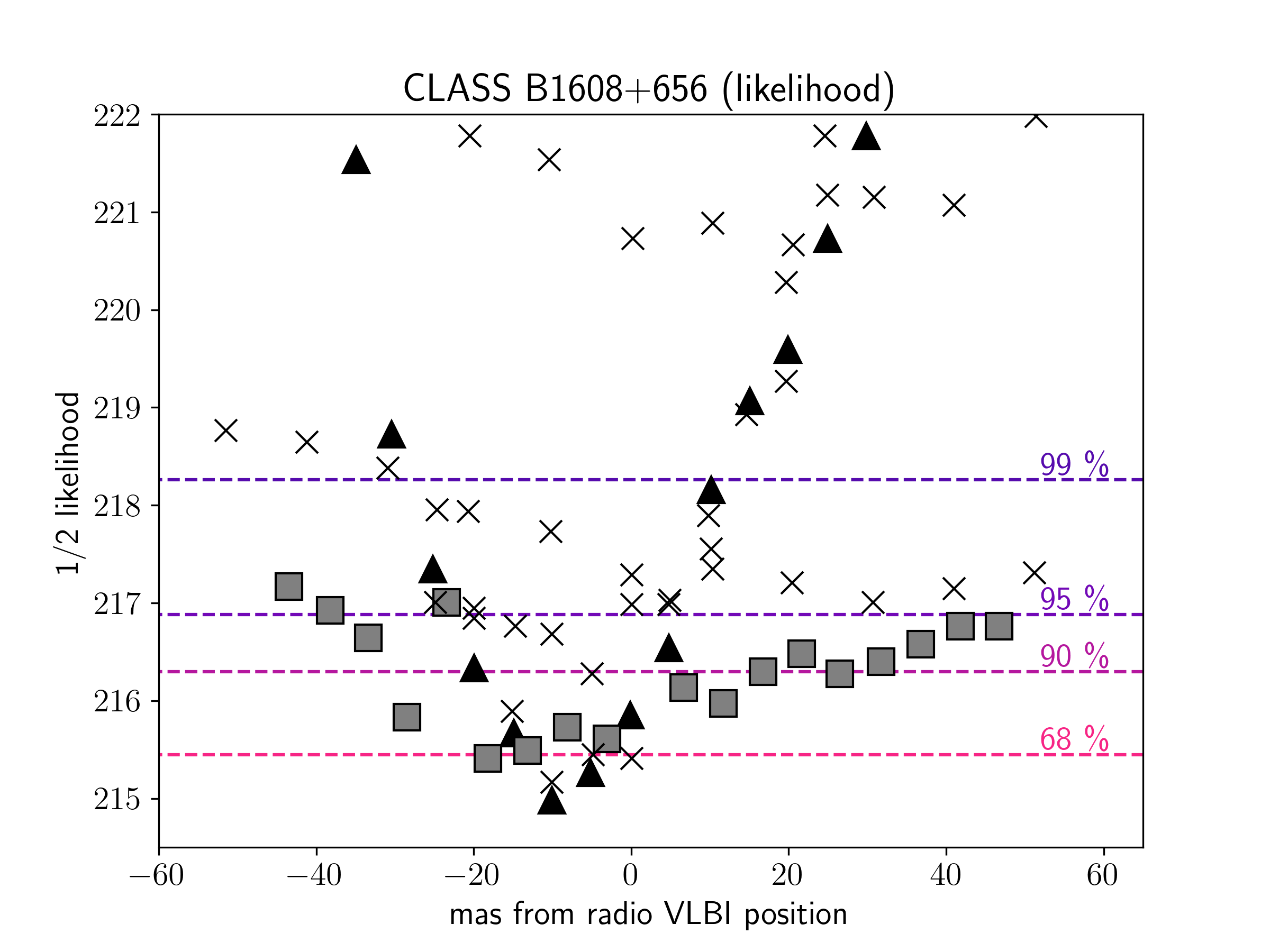} \\
        \includegraphics[width=0.7\textwidth]{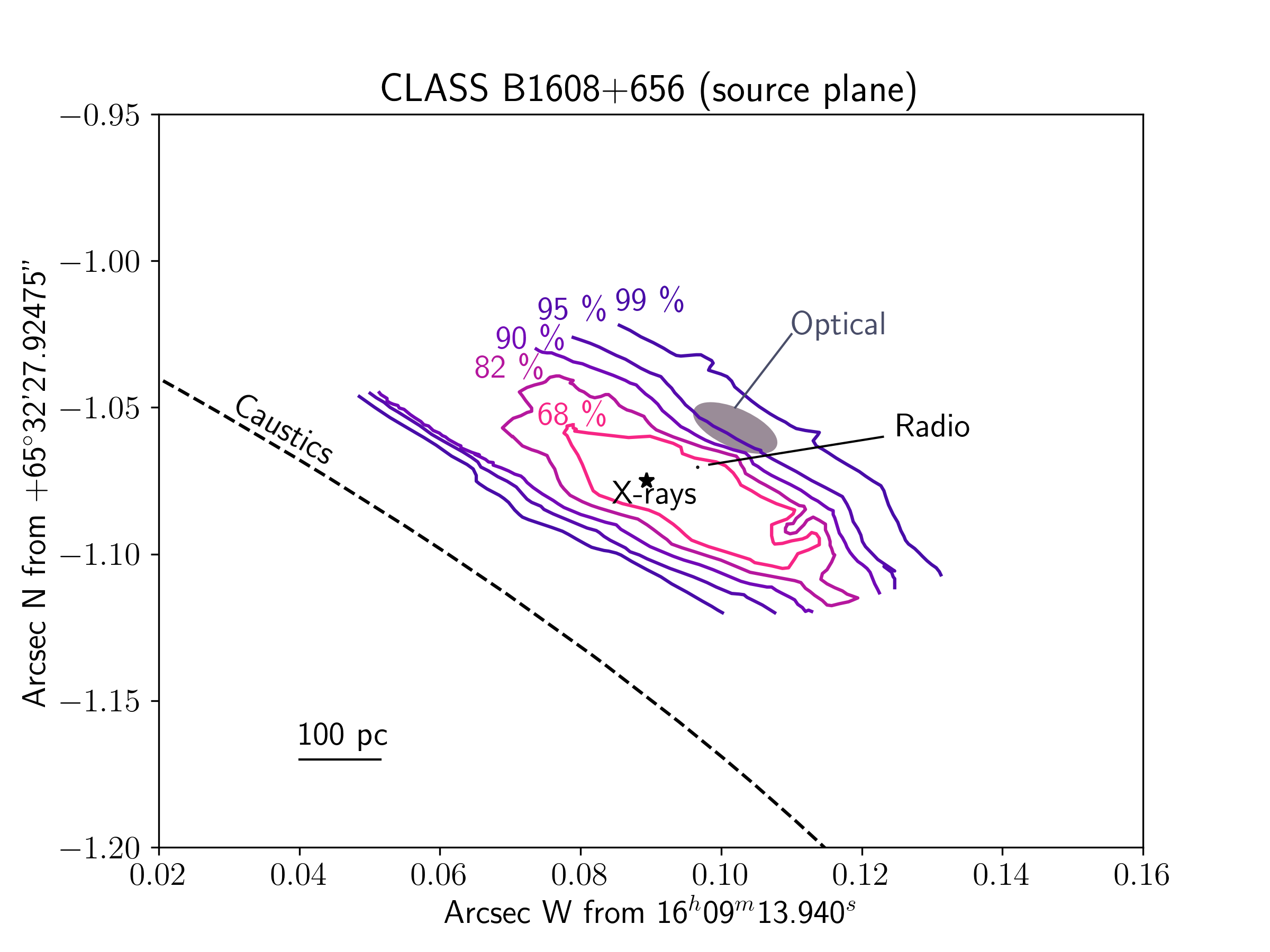}
     \caption{\textsl{Top}: Likelihood values for different trial source positions of CLASS B1608+656. Black are perpendicular to caustic, grey are parallel to caustic. The squares and triangles are the points shown in Figure~\ref{fig:b0712_simulated_sources}, and the black crosses are additional lines of points chosen to define the contours. \textsl{Bottom}: Contours of allowed X-ray source plane position. The three outer contours, 99\%, 95\%, and 90\%, respectively, do not close parallel to the caustic, due to limited X-ray statistics and the very small change in the potential gradient in that direction. The inner two contours are at 82\% and 68\% confidence. The black and grey ellipses indicate the radio VLBI and the HST optical source positions, respectively.  The size of the ellipses indicate the 1$\sigma$ uncertainty. The black dashed line shows a portion of the caustic. We highlight that the elongation of the contours is caused by the almost constant magnification factor in the direction parallel to the caustic line. 
    }
    \label{fig:source_reconstruction_b1608}
\end{figure*}

Figures \ref{fig:source_reconstruction_b0712} and \ref{fig:source_reconstruction_b1608} show the X-ray source localization for CLASS B0712+472 and CLASS B1608+656, respectively.

For CLASS B0712+472, the X-ray source is at (-8.5, +6.3) mas with respect to the VLBI source plane position.
The best position has 1$\sigma$ error bars $_{-2}^{+3}$ mas perpendicular to the caustic, (where the minus is to the SE and the plus to the NW) and $-14$ mas (to the SW) $+11$ mas (to the NE) parallel to the caustic. The measured X-ray source position is about 1$\sigma$ offset from the radio VLBI source, and offset, at 90\% confidence, with respect to the optical emission (Fig. \ref{fig:source_reconstruction_b0712}). \citet{Spingola&Barnacka2020} measured the radio and optical emissions to be within (2 $\pm$ 5) mas, so the present result allows the X-rays to also be co-spatial with both those bands.

The X-ray emission in CLASS B1608+656 is located ($-7.2$, $-4.8$) mas relative to the VLBI source plane position, with 1$\sigma$ errors ($+11$ to NW, $-8$ to SE) mas perpendicular to caustic, and ($-35$ to SW , $+20$ to NE) mas parallel to caustic (Fig. \ref{fig:source_reconstruction_b1608}).
The parallel results do not converge to closed contours beyond 82\% confidence, since with 10 times fewer counts we are much less sensitive than for CLASS~B0712+472 (Fig. \ref{fig:source_reconstruction_b1608}). Nevertheless, such astrometric precision is remarkable for X-ray observations at $z=1.394$ (we discuss this in Sec.~\ref{sec:discussion}). Also in this case, within the positional uncertainties the X-ray source is co-spatial with the radio emissions, while the optical source uncertainty extends within the 90\% confidence location of the X-ray source. 

In both sources there is an evident elongation of the contours in the direction parallel to the caustic (Figs. \ref{fig:source_reconstruction_b0712} and \ref{fig:source_reconstruction_b1608}). We remind the reader that the magnification changes in the direction perpendicular to the caustics and its quite steep close to the caustics. Instead, it is pretty constant in the direction parallel to the caustics \citep{Barnacka2017}. Therefore, it is natural that the better constraints on the X-ray source location are in the direction perpendicular to the caustics.

\section{Discussion}\label{sec:discussion}

\subsection{Importance of astrometric precision}

Astrometry is essential to all areas of astrophysics. For instance, the precise measurement of positions allows astronomers to study the structure and formation of our Galaxy \citep[e.g.,][]{Helmi2018, Belokurov2020, Libralato2021}, determine the dark matter profile in galaxies in the Local Group \citep[e.g.,][]{Massari2018, Massari2020}, associate sources observed at different wavelengths \citep[e.g.,][]{Dabrusco2019, Lindegren2020}, study the density profile of lensing galaxies on sub-galactic scales \citep[e.g.,][]{Chen2007, Sluse2012,Spingola2018}, measure AGN jets proper motions at cosmological distances \citep[e.g.,][]{Frey2015, Perger2018, An2020}.  A precise localization of the radio/X-ray emission from the AGN and the peak of optical emission of their host galaxy is crucial to identify and confirm dual and offset AGN, especially those at small angular separation, which trace the final stages of galaxy merging \citep[e.g.,][]{Deane2014}. To fully characterize these last phases, it is necessary to reach a pc-scale astrometric precision also at cosmological epochs, when mergers were more common. At X-rays this is particularly difficult, as at $z=1$ the spatial resolution of \axaf\ is of $\sim 4$ kpc, not comparable to the spatial resolution that can be achieved, for example, with VLBI at high redshifts \citep[e.g.,][]{Spingola2020, Perger2021, Momjian2021, Zhang2021}. Also, the astrometric precision depends on the significance of the source detection (i.e., signal-to-noise ratio). Therefore, reaching a sub-kpc astrometric precision is possible only for the brightest sources, which may not be representative of the entire AGN population at high-$z$ \citep[e.g.,][]{Bosco2021}. For these reasons, to date at high redshift only binary/offset SMBHs at wide angular separation have been found, which trace the very early stages of galaxy merging \citep[e.g.,][]{Vito2021}.

\subsection{Milliarcsecond astrometric precision at X-rays}

In this work, we exploit non-linear amplification of the lens caustics combined with a novel Bayesian method to obtain precise measurement of the X-ray emission to search for binary/offset AGN systems at a small angular separation at $z>1$. The caustic provides reference frame and its ability to amplify sources changes with the source location, projection and distance to the caustic. In one dimension, perpendicular to the caustic, we obtain X-ray astrometric precision (1$\sigma$) of order 2.5~mas ($\sim20$~pc, with 845 photons for CLASS B0712+472) and 10~mas ($\sim80$~pc, with 79 photons for CLASS B1608+656). Such precision allows us to confirm that the AGN in CLASS B0712+472 is located at the center of its host galaxy (traced by the peak of the optical emission, \citealt{Spingola&Barnacka2020}), while the X-ray emission in CLASS B1608+656 is offset with respect to the optical at 1$\sigma$ level, but co-spatial within 3$\sigma$ (Fig.~\ref{fig:source_reconstruction_b1608}). The lower number of photons detected for CLASS B1608+656 results in a worse astrometric precision, which prevents us to securely confirm the offset AGN nature of this object. 

In both cases we reach an astrometric precision that at X-rays is comparable only to local X-ray observations (i.e., tens of milliarcseconds,  \citealt{Kong2010, Ratti2010, Auchettl2015, Ponti2019, Tomsick2020, Tomsick2021}). This precise location is what we are looking for identify binary/offset AGN candidates at a small physical separation in the early Universe (see Sec.~\ref{sec:lensed_offset_dual_AGN}). 
The method allows us to improve on the absolute astrometry of \axaf\ instrument by two orders of magnitude. 
It terms of relative astrometry, the method gives us a factor of 10 by allowing us to translate accuracy of relative astrometry of dozens of milliarseconds in the lens image plane to a few milliseconds accuracy in the source plane. 
Our ability to improve on the astrometry is given by how well we can measure relative positions of the lensed images of the source, which depends on the photon statistics. The second key factor is the distance of the source from the caustic. The sources located closer to the caustic of the lens experience higher flux magnification and relative positions of the lensed images change more drastically. Both sources, CLASS B0712+472 and CLASS B1608+656, are at a moderate distance from the caustic (8 and 37 mas, respectively), and are magnified by a factor of $\sim 10$ ($8$ and $14$, respectively).  The two sources are, therefore, located from the caustic curve at 1.3\% of the Einstein radius (E.R.) in CLASS B0712+472 and 4.1\% E.R. in CLASS B1608+656. As demonstrated by \citealt{Barnacka2017}, the most significant amplification of any offsets between the multi-band emission can be observed when the source is within 2.0\% E.R. to the caustic. Although this is an extremely rare lensing configuration, less than 1\%  probability, the magnification bias increases the probability of observing those sources by an order of magnitude (\citealt{Barnacka2017}).

Previously, we applied this method to the X-ray observations of the radio-loud lensing system MG~B2016+112 \citep{Schwartz2021}. From MG~B2016+112 we detected only 24 photons, which we found to come from two X-ray sources, making this object a promising dual AGN candidate at $z=3.273$ (\citealt{Schwartz2021}, but see also \citealt{Spingola2019}). We could locate the two X-ray sources at a precision of 100~mas, which correspond to $\sim 800$~pc at the redshift of the source.

We highlight that in all these cases the lens mass models are precise, having uncertainties on the set of mass parameters smaller than 10\% \citep{Spingola2019, Spingola&Barnacka2020}. Such precision in the lens mass models can be achieved only with stringent constraints, which are primarily given by position and fluxes of the lensed images. Therefore, combining VLBI- and optical-derived models with X-ray observations is a powerful way to locate the X-ray emission at pc-scale precision at high redshifts. Including MG~B2016+112 in our sample of three lensed AGN at $z>1$, our method spatially locates the X-ray emission in two dimensions at a precision between 25 and 800~parsecs. \\

\begin{figure*}
    \centering
    \includegraphics[width =1.0\textwidth]{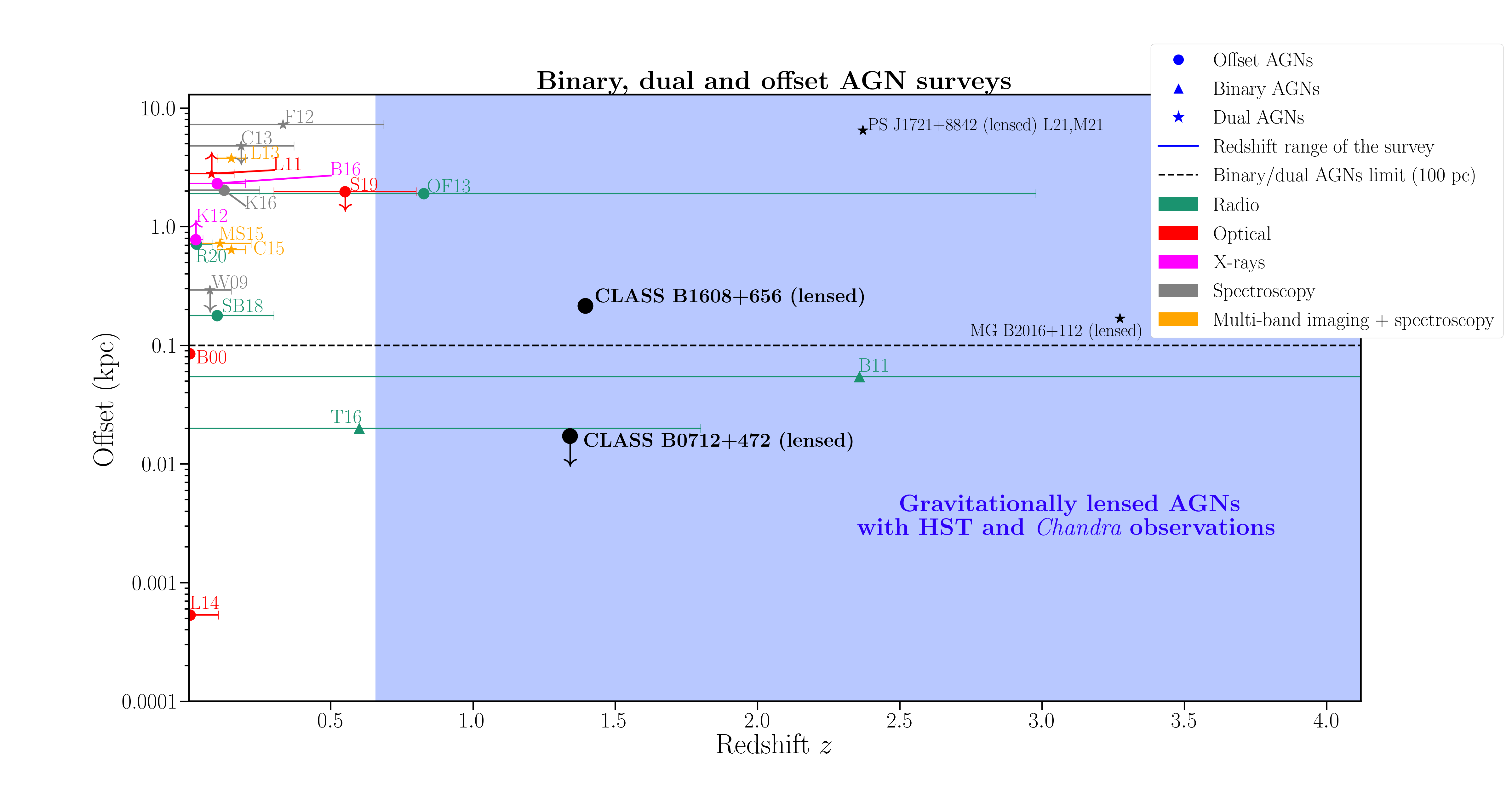}
    \caption{ \small Offset (kpc, logarithmic scale) as a function of redshift for the largest binary, dual and offset AGN searches. With \textsl{offset} we mean the separation between the optical peak of the emission of the host galaxy and radio/X-ray emissions (for offset AGN) or the angular separation between the two AGN (for binary AGN).
    The surveys are colored by observing bands (as indicated in the legend on the top right). The average offset found in the offset AGN surveys is shown with a filled circle (\citealt{Binggeli2000}, B00; \citealt{Koss2012}, K12;  \citealt{Orosz2013}, OF13; \citealt{Lena2014}, L14; \citealt{Kim2016}, K16; \citealt{Barrows2016}, B16; \citealt{Skipper2018}, SB18; \citealt{Shen2019}, S19;  \citealt{Reines2020}, R20), while the average separation of two AGN in binary AGN surveys is shown using a filled triangle (\citealt{Wang2009}, W09; \citealt{Burke-Spolaor2011}, B11; \citealt{Liu2011}, L11; \citealt{Fu2012}, F12; \citealt{Comerford2013}, C13; \citealt{Comerford2015}, C15; \citealt{Muller-Sanchez2015}, MS15; \citealt{Tremblay2016}, T16). Upper and lower limits on the offsets are shown with up/down arrows. The black dashed line indicates the canonical separation between binary and dual AGN (100 pc), which is roughly the Bondi radius of a SMBH of $5\times10^8$ M$_{\odot}$. We plot in black our results for the three lensing systems CLASS~B0712+472, CLASS~B1608+656 (this work and \citealt{Spingola&Barnacka2020}) and MG~B2016+112 \citep{Spingola2019, Schwartz2021}, which cover the smallest angular separations and highest redshifts. We also plot with a black triangle the recently discovered lensed dual AGN PS J1721+8842 \citep{Lemon2021, Mangat2021}. We plan to apply our method to the entire \axaf\ lensing sample, which covers a redshift range $z=0.66-4.12$ (blue area).}
    \label{fig:redshift_range}
\end{figure*}

\subsection{Gravitationally lensed offset and binary AGN}
\label{sec:lensed_offset_dual_AGN}

Gravitationally lensed sources provide an ideal unbiased sample to assess the frequency of binary, dual and offset AGN at high redshift, which is crucial to test SMBHs formation and evolution models. For example, finding even one lensed dual AGN implies a higher fraction of such systems with respect to what current simulations predict, making dual AGN systems potentially up to an order of magnitude more abundant than currently thought \citep[e.g.,][]{Spingola2019, Rosas-guevara2019}. This is reinforced by the recent discovery at optical wavelengths of the lensed dual AGN PS J1721+8812 at $z\sim 2.38$ \citep{Lemon2021}. The two lensed AGN are at a wider separation ($\sim6$ kpc) than MG B2016+112 ($\sim175$ pc), as also confirmed by radio observations \citep{Mangat2021}. Another lensed binary AGN candidate is the blazar PKS 1830$-$211 at $z=2.51$: \citet{Nair2005} suggested that the measured precession period of the jet could be also consistent with the system hosting two SMBHs.
 
A more systematic approach is needed to better assess the rate of binary and offset AGN at high redshifts. In particular, quadruply imaged sources are at a premium for these studies, because their higher magnification allows us to measure smaller offsets \citep{Barnacka2017, Barnacka2018}. In fact, in the two quadruply imaged systems analyzed in this work we could spatially locate the X-ray emission on a scale of hundreds of parsecs in projection. This is an excellent precision to discover multiple and offset AGN that are undergoing the final stages of galaxy merging. Inferring the percentage of offset and binary AGN in the early Universe is a powerful probe of the duration of the inspiral timescale and its connection to the circumbinary disk \citep[e.g.,][]{Lodato2009, Rafikov2016, LopezArmengol2021}.

In Fig. \ref{fig:redshift_range} we show that currently our method can find the smallest offset and binary AGN systems at $z>1$. We compare our results with the largest offset, binary  and dual AGN searches, which use several observing bands. In order to find offset AGN typically one measures the distance between the emission from the SMBH (X-ray, radio or optical point-like emission) from the peak of the optical emission of the host galaxy \citep{Binggeli2000, Koss2012, Orosz2013, Lena2014, Kim2016, Barrows2016, Skipper2018, Shen2019, Reines2020}. Instead, the presence of two flat-spectrum radio sources is generally a strong indication for pairs of AGN, as also the presence of double peaks in the narrow emission lines \citep{Wang2009, Burke-Spolaor2011, Liu2011, Fu2012, Comerford2013, Comerford2015, Muller-Sanchez2015, Tremblay2016}. All these methods can robustly find offset and binary AGN, but they are mostly limited to the local Universe ($z<1$) because of the necessary sensitivity and angular resolution to reveal such systems. Unveiling offset and pairs of non-lensed AGN at high redshift at the smallest separations (i.e., $<1$ kpc) can be currently done only by using radio VLBI observations (\citealt{Burke-Spolaor2011,Orosz2013, Tremblay2016}, Fig. \ref{fig:redshift_range}). However, radio-loud AGN are rare, as only up to 10 \% of the total AGN population is radio-loud \citep{Padovani2017}, thus they trace only a small portion of all SMBHs.

\subsection{Intrinsic X-ray properties of two lensed AGN}

The X-ray emission from AGN might come from the innermost regions close to the SMBH (the accretion disc-corona; e.g., \citealt{Begelman1983,Haardt1991}) and, if present, the jets \citep[e.g.,][]{Chartas2000_jet,Harris2003, Worrall2009, Blandford2019}. Assessing the origin of the X-ray emission is important to firmly classify a system an offset or dual AGN. For instance, a dual AGN candidate could actually consist of a core-jet AGN, and an observed offset between the optical and X-ray emission can be attributed to an unresolved extended jet which is offset from the central optical quasar-like emission. The X-ray emission in jets is complex, as it involves several mechanisms  \citep[e.g.,][]{Schwartz2000} and can be extended on several kpc  \citep[e.g.,][]{Schwartz2020}. At high redshift the studies of the X-ray properties of AGN are mostly limited to the brightest ($>10^{44}$ erg s$^{-1}$) sources \citep[e.g.,][]{Vito2019, Snios2021, Ighina2021}.

There is no evidence for extended X-ray emission in the two lensed AGN analysed in this work. Therefore, we consider the emission coming from a compact point-like source at the position of our best localization (Figs. \ref{fig:source_reconstruction_b0712} and \ref{fig:source_reconstruction_b1608}), where the magnification factor is $\mu_{\rm B0712}^{X-ray} = 22\pm9$ and $\mu_{\rm B1608}^{X-ray} = 15\pm1$ for CLASS B0712+472 and CLASS B1608+656, respectively.
We obtain an intrinsic (de-lensed)  luminosity in the 0.5 to 7 keV rest-frame of $1.9\pm 0.7 \times 10 ^{43}$ erg s$^{-1}$ for CLASS B0712+472 and $1.5 \pm 0.4 \times 10 ^{43}$ erg s$^{-1}$ for CLASS B1608+656. Therefore, these two AGN have an X-ray luminosity at soft energies that places them neither to the brightest nor to the faintest AGN \citep[e.g.,][]{Gilli2009, Aird2015}, with the caveat that these values come from a single-epoch observation and, therefore, variability can play a role.  

We found X-ray indices that are a few tenths flatter than typical (Sec. \ref{sec:observations}), which can be attributed to intrinsic absorption at the source redshift with a column density $N_H =   (2.5\pm2.2)\times10^{22}$ cm$^{-2}$ in CLASS B0712+472 and $N_H< 2 \times 10^{21}$ cm$^{-2}$  in CLASS B1608+656. Therefore, these two sources have low intrinsic absorption and can be considered Compton thin AGN.  Similar values have been found in other the multiple imaged AGN over a wide range of redshifts, such as PG 1115+080 ($z=1.72$, \citealt{Chartas2000, Chartas2007}) and APM 08279+5255 ($z=3.91$, \citealt{Chartas2002_apm}), MG J0414+0534 ($z=2.64$, \citealt{Chartas2002}) and MG B2016+112 ($z=3.273$, \citealt{Chartas2001, Schwartz2021}).

\section{Conclusions and future perspectives} \label{sec:conclusions}

In this paper we presented the localization of the X-ray emission detected by \textsl{Chandra} in two gravitationally lensed sources CLASS B0712+472 and CLASS B1608+656 for the first time. Thanks to precise lens mass models and our novel Bayesian technique, we could infer the source position at milliarcsecond precision. The X-ray sources in CLASS B0712+472 and CLASS B1608+656 are located within $2\pm4$ mas and $8^{+29}_{-17}$ mas in projection from the radio source, respectively. This is an unprecedented astrometric precision for X-ray observations of sources at redshifts 1.34 (CLASS B0712+472, 845 photons detected) and 1.394 (CLASS B1608+656, 79 photons detected). We found that the optical, radio and X-ray emissions are co-spatial within uncertainties in CLASS B0712+472. Instead, the optical emission is offset at 1$\sigma$ level with respect to the X-ray and radio emission in CLASS B1608+656, which is a promising offset AGN candidate.

These results demonstrate how gravitational lensing can unveil multi-wavelength offsets of tens of parsecs in a redshift range that is most critical to test galaxy evolution models, and unaccessible with current instruments otherwise. This method can, therefore, be used to find offset and binary AGN, which are the primary target sources of the future missions of LISA and PTA.

We plan to apply our novel technique to the entire sample of X-ray-loud strongly lensed AGN, which covers the redshift range $z=0.66-4.12$ (Fig. \ref{fig:redshift_range}). Moreover, some of the lensed AGN have been detected by \textsl{Gaia}, which can provide $\mu$as astrometry for the brightest sources \citep{Brown2021}. In the foreseeable future, surveys with the ``Vera C. Rubin" Observatory and the Square Kilometer Array will find $\sim10^5$ lenses \citep{Collett2015, McKean2015}. Among these sources there will be offset and pairs of AGN, which will provide a statistically significant sample to finally assess the fraction of these systems at high redshift. \\

\section*{Acknowledgement}

The authors would like to thank the anonymous referee for the constructive comments on this manuscript.

CS acknowledges financial support from the Italian Ministry of University and Research - Project Proposal CIR01\_00010.
This work has been supported by NASA contract NAS8-03060
to SAO, and grant GO8-19077X from the CXC. This research made use of
the NASA Astrophysics Data System.  The authors thank Alysa Rogers for her help with the background source simulations.

\facilities{\axaf.}

\software{ciao-4.12 \citep{Fruscione2006}, SAOTrace-2.0.4\_03 \citep{Jerius2004}, Marx-5.5.0 \citep{Davis2012}, SAOImageDS9 Version 8.2b1 \citep{Joye2003}, Gravlens \citep{Keeton2001a, Keeton2001b}, APLpy \citep{Robitaille2012}.}

\bibliography{references}{}

\begin{thebibliography}{}
\expandafter\ifx\csname natexlab\endcsname\relax\def\natexlab#1{#1}\fi
\providecommand{\url}[1]{\href{#1}{#1}}
\providecommand{\dodoi}[1]{doi:~\href{http://doi.org/#1}{\nolinkurl{#1}}}
\providecommand{\doeprint}[1]{\href{http://ascl.net/#1}{\nolinkurl{http://ascl.net/#1}}}
\providecommand{\doarXiv}[1]{\href{https://arxiv.org/abs/#1}{\nolinkurl{https://arxiv.org/abs/#1}}}

\bibitem[{{Aird} {et~al.}(2015){Aird}, {Coil}, {Georgakakis}, {Nandra},
  {Barro}, \& {P{\'e}rez-Gonz{\'a}lez}}]{Aird2015}
{Aird}, J., {Coil}, A.~L., {Georgakakis}, A., {et~al.} 2015, \mnras, 451, 1892,
  \dodoi{10.1093/mnras/stv1062}

\bibitem[{{An} {et~al.}(2020){An}, {Mohan}, {Zhang}, {Frey}, {Yang},
  {Gab{\'a}nyi}, {Gurvits}, {Paragi}, {Perger}, \& {Zheng}}]{An2020}
{An}, T., {Mohan}, P., {Zhang}, Y., {et~al.} 2020, Nature Communications, 11,
  143, \dodoi{10.1038/s41467-019-14093-2}

\bibitem[{{Auchettl} {et~al.}(2015){Auchettl}, {Slane}, {Romani}, {Posselt},
  {Pavlov}, {Kargaltsev}, {Ng}, {Temim}, {Weisskopf}, {Bykov}, \&
  {Swartz}}]{Auchettl2015}
{Auchettl}, K., {Slane}, P., {Romani}, R.~W., {et~al.} 2015, \apj, 802, 68,
  \dodoi{10.1088/0004-637X/802/1/68}

\bibitem[{{Auger} {et~al.}(2009){Auger}, {Treu}, {Bolton}, {Gavazzi},
  {Koopmans}, {Marshall}, {Bundy}, \& {Moustakas}}]{Auger2009}
{Auger}, M.~W., {Treu}, T., {Bolton}, A.~S., {et~al.} 2009, \apj, 705, 1099,
  \dodoi{10.1088/0004-637X/705/2/1099}

\bibitem[{{Barnacka}(2017)}]{Barnacka2017}
{Barnacka}, A. 2017, \apj, 846, 157, \dodoi{10.3847/1538-4357/aa86ec}

\bibitem[{{Barnacka}(2018)}]{Barnacka2018}
---. 2018, \physrep, 778, 1, \dodoi{10.1016/j.physrep.2018.10.001}

\bibitem[{{Barnacka} {et~al.}(2015){Barnacka}, {Geller}, {Dell'Antonio}, \&
  {Benbow}}]{Barnacka2015}
{Barnacka}, A., {Geller}, M.~J., {Dell'Antonio}, I.~P., \& {Benbow}, W. 2015,
  \apj, 809, 100, \dodoi{10.1088/0004-637X/809/1/100}

\bibitem[{{Barnacka} {et~al.}(2016){Barnacka}, {Geller}, {Dell'Antonio}, \&
  {Zitrin}}]{Barnacka2016}
{Barnacka}, A., {Geller}, M.~J., {Dell'Antonio}, I.~P., \& {Zitrin}, A. 2016,
  \apj, 821, 58, \dodoi{10.3847/0004-637X/821/1/58}

\bibitem[{{Barrows} {et~al.}(2018){Barrows}, {Comerford}, \&
  {Greene}}]{Barrows2018}
{Barrows}, R.~S., {Comerford}, J.~M., \& {Greene}, J.~E. 2018, \apj, 869, 154,
  \dodoi{10.3847/1538-4357/aaedb6}

\bibitem[{{Barrows} {et~al.}(2016){Barrows}, {Comerford}, {Greene}, \&
  {Pooley}}]{Barrows2016}
{Barrows}, R.~S., {Comerford}, J.~M., {Greene}, J.~E., \& {Pooley}, D. 2016,
  \apj, 829, 37, \dodoi{10.3847/0004-637X/829/1/37}

\bibitem[{{Bartlett} {et~al.}(2021){Bartlett}, {Desmond}, {Devriendt},
  {Ferreira}, \& {Slyz}}]{Bartlett2021}
{Bartlett}, D.~J., {Desmond}, H., {Devriendt}, J., {Ferreira}, P.~G., \&
  {Slyz}, A. 2021, \mnras, 500, 4639, \dodoi{10.1093/mnras/staa3516}

\bibitem[{{Begelman} \& {McKee}(1983)}]{Begelman1983}
{Begelman}, M.~C., \& {McKee}, C.~F. 1983, \apj, 271, 89,
  \dodoi{10.1086/161179}

\bibitem[{{Belokurov} {et~al.}(2020){Belokurov}, {Sanders}, {Fattahi}, {Smith},
  {Deason}, {Evans}, \& {Grand}}]{Belokurov2020}
{Belokurov}, V., {Sanders}, J.~L., {Fattahi}, A., {et~al.} 2020, \mnras, 494,
  3880, \dodoi{10.1093/mnras/staa876}

\bibitem[{{Berta} {et~al.}(2021){Berta}, {Young}, {Cox}, {Neri}, {Jones},
  {Baker}, {Omont}, {Dunne}, {Carnero Rosell}, {Marchetti}, {Negrello}, {Yang},
  {Riechers}, {Dannerbauer}, {Perez-Fournon}, {van der Werf}, {Bakx}, {Ivison},
  {Beelen}, {Buat}, {Cooray}, {Cortzen}, {Dye}, {Eales}, {Gavazzi}, {Harris},
  {Herrera}, {Hughes}, {Jin}, {Krips}, {Lagache}, {Lehnert}, {Messias},
  {Serjeant}, {Stanley}, {Urquhart}, {Vlahakis}, \& {Wei{\ss}}}]{Berta2021}
{Berta}, S., {Young}, A.~J., {Cox}, P., {et~al.} 2021, \aap, 646, A122,
  \dodoi{10.1051/0004-6361/202039743}

\bibitem[{{Binggeli} {et~al.}(2000){Binggeli}, {Barazza}, \&
  {Jerjen}}]{Binggeli2000}
{Binggeli}, B., {Barazza}, F., \& {Jerjen}, H. 2000, \aap, 359, 447

\bibitem[{{Blandford} {et~al.}(2019){Blandford}, {Meier}, \&
  {Readhead}}]{Blandford2019}
{Blandford}, R., {Meier}, D., \& {Readhead}, A. 2019, \araa, 57, 467,
  \dodoi{10.1146/annurev-astro-081817-051948}

\bibitem[{{Bosco} {et~al.}(2021){Bosco}, {Hennawi}, {Stern}, \&
  {Pott}}]{Bosco2021}
{Bosco}, F., {Hennawi}, J.~F., {Stern}, J., \& {Pott}, J.-U. 2021, arXiv
  e-prints, arXiv:2106.15900.
\newblock \doarXiv{2106.15900}

\bibitem[{{Brown}(2021)}]{Brown2021}
{Brown}, A. G.~A. 2021, arXiv e-prints, arXiv:2102.11712.
\newblock \doarXiv{2102.11712}

\bibitem[{{Burke-Spolaor}(2011)}]{Burke-Spolaor2011}
{Burke-Spolaor}, S. 2011, \mnras, 410, 2113,
  \dodoi{10.1111/j.1365-2966.2010.17586.x}

\bibitem[{{Burke-Spolaor} {et~al.}(2014){Burke-Spolaor}, {Brazier},
  {Chatterjee}, {Comerford}, {Cordes}, {Lazio}, {Liu}, \&
  {Shen}}]{Burke-Spolaor2014}
{Burke-Spolaor}, S., {Brazier}, A., {Chatterjee}, S., {et~al.} 2014, arXiv
  e-prints, arXiv:1402.0548.
\newblock \doarXiv{1402.0548}

\bibitem[{{Burke-Spolaor} {et~al.}(2019){Burke-Spolaor}, {Taylor}, {Charisi},
  {Dolch}, {Hazboun}, {Holgado}, {Kelley}, {Lazio}, {Madison}, {McMann},
  {Mingarelli}, {Rasskazov}, {Siemens}, {Simon}, \&
  {Smith}}]{Burke-Spolaor2019}
{Burke-Spolaor}, S., {Taylor}, S.~R., {Charisi}, M., {et~al.} 2019, \aapr, 27,
  5, \dodoi{10.1007/s00159-019-0115-7}

\bibitem[{{Cash}(1979)}]{Cash1979}
{Cash}, W. 1979, \apj, 228, 939, \dodoi{10.1086/156922}

\bibitem[{{Chartas}(2000)}]{Chartas2000}
{Chartas}, G. 2000, \apj, 531, 81, \dodoi{10.1086/308441}

\bibitem[{{Chartas} {et~al.}(2002{\natexlab{a}}){Chartas}, {Agol}, {Eracleous},
  {Garmire}, {Bautz}, \& {Morgan}}]{Chartas2002}
{Chartas}, G., {Agol}, E., {Eracleous}, M., {et~al.} 2002{\natexlab{a}}, \apj,
  568, 509, \dodoi{10.1086/339162}

\bibitem[{{Chartas} {et~al.}(2001){Chartas}, {Bautz}, {Garmire}, {Jones}, \&
  {Schneider}}]{Chartas2001}
{Chartas}, G., {Bautz}, M., {Garmire}, G., {Jones}, C., \& {Schneider}, D.~P.
  2001, \apjl, 550, L163, \dodoi{10.1086/319632}

\bibitem[{{Chartas} {et~al.}(2002{\natexlab{b}}){Chartas}, {Brandt},
  {Gallagher}, \& {Garmire}}]{Chartas2002_apm}
{Chartas}, G., {Brandt}, W.~N., {Gallagher}, S.~C., \& {Garmire}, G.~P.
  2002{\natexlab{b}}, \apj, 579, 169, \dodoi{10.1086/342744}

\bibitem[{{Chartas} {et~al.}(2007){Chartas}, {Brandt}, {Gallagher}, \&
  {Proga}}]{Chartas2007}
{Chartas}, G., {Brandt}, W.~N., {Gallagher}, S.~C., \& {Proga}, D. 2007, \aj,
  133, 1849, \dodoi{10.1086/512364}

\bibitem[{{Chartas} {et~al.}(2000){Chartas}, {Worrall}, {Birkinshaw},
  {Cresitello-Dittmar}, {Cui}, {Ghosh}, {Harris}, {Hooper}, {Jauncey}, {Kim},
  {Lovell}, {Mathur}, {Schwartz}, {Tingay}, {Virani}, \&
  {Wilkes}}]{Chartas2000_jet}
{Chartas}, G., {Worrall}, D.~M., {Birkinshaw}, M., {et~al.} 2000, \apj, 542,
  655, \dodoi{10.1086/317049}

\bibitem[{{Chen} {et~al.}(2007){Chen}, {Rozo}, {Dalal}, \& {Taylor}}]{Chen2007}
{Chen}, J., {Rozo}, E., {Dalal}, N., \& {Taylor}, J.~E. 2007, \apj, 659, 52,
  \dodoi{10.1086/512002}

\bibitem[{{Collett}(2015)}]{Collett2015}
{Collett}, T.~E. 2015, \apj, 811, 20, \dodoi{10.1088/0004-637X/811/1/20}

\bibitem[{{Comerford} {et~al.}(2012){Comerford}, {Gerke}, {Stern}, {Cooper},
  {Weiner}, {Newman}, {Madsen}, \& {Barrows}}]{Comerford2012}
{Comerford}, J.~M., {Gerke}, B.~F., {Stern}, D., {et~al.} 2012, \apj, 753, 42,
  \dodoi{10.1088/0004-637X/753/1/42}

\bibitem[{{Comerford} {et~al.}(2015){Comerford}, {Pooley}, {Barrows}, {Greene},
  {Zakamska}, {Madejski}, \& {Cooper}}]{Comerford2015}
{Comerford}, J.~M., {Pooley}, D., {Barrows}, R.~S., {et~al.} 2015, \apj, 806,
  219, \dodoi{10.1088/0004-637X/806/2/219}

\bibitem[{{Comerford} {et~al.}(2013){Comerford}, {Schluns}, {Greene}, \&
  {Cool}}]{Comerford2013}
{Comerford}, J.~M., {Schluns}, K., {Greene}, J.~E., \& {Cool}, R.~J. 2013,
  \apj, 777, 64, \dodoi{10.1088/0004-637X/777/1/64}

\bibitem[{{Congdon} \& {Keeton}(2018)}]{Congdon2018}
{Congdon}, A.~B., \& {Keeton}, C. 2018, {Principles of Gravitational Lensing:
  Light Deflection as a Probe of Astrophysics and Cosmology}

\bibitem[{{Conselice} {et~al.}(2003){Conselice}, {Bershady}, {Dickinson}, \&
  {Papovich}}]{Conselice2003}
{Conselice}, C.~J., {Bershady}, M.~A., {Dickinson}, M., \& {Papovich}, C. 2003,
  \aj, 126, 1183, \dodoi{10.1086/377318}

\bibitem[{{D'Abrusco} {et~al.}(2019){D'Abrusco}, {{\'A}lvarez Crespo},
  {Massaro}, {Campana}, {Chavushyan}, {Landoni}, {La Franca}, {Masetti},
  {Milisavljevic}, {Paggi}, {Ricci}, \& {Smith}}]{Dabrusco2019}
{D'Abrusco}, R., {{\'A}lvarez Crespo}, N., {Massaro}, F., {et~al.} 2019, \apjs,
  242, 4, \dodoi{10.3847/1538-4365/ab16f4}

\bibitem[{{Dai} \& {Kochanek}(2005)}]{Dai2005}
{Dai}, X., \& {Kochanek}, C.~S. 2005, \apj, 625, 633, \dodoi{10.1086/429485}

\bibitem[{{Dai} \& {Kochanek}(2009)}]{Dai2009}
---. 2009, \apj, 692, 677, \dodoi{10.1088/0004-637X/692/1/677}

\bibitem[{{Davis} {et~al.}(2012){Davis}, {Bautz}, {Dewey}, {Heilmann}, {Houck},
  {Huenemoerder}, {Marshall}, {Nowak}, {Schattenburg}, {Schulz}, \&
  {Smith}}]{Davis2012}
{Davis}, J.~E., {Bautz}, M.~W., {Dewey}, D., {et~al.} 2012, in Society of
  Photo-Optical Instrumentation Engineers (SPIE) Conference Series, Vol. 8443,
  Space Telescopes and Instrumentation 2012: Ultraviolet to Gamma Ray, ed.
  T.~{Takahashi}, S.~S. {Murray}, \& J.-W.~A. {den Herder}, 84431A,
  \dodoi{10.1117/12.926937}

\bibitem[{{De Rosa} {et~al.}(2019){De Rosa}, {Vignali}, {Bogdanovi{\'c}},
  {Capelo}, {Charisi}, {Dotti}, {Husemann}, {Lusso}, {Mayer}, {Paragi},
  {Runnoe}, {Sesana}, {Steinborn}, {Bianchi}, {Colpi}, {del Valle}, {Frey},
  {Gab{\'a}nyi}, {Giustini}, {Guainazzi}, {Haiman}, {Herrera Ruiz},
  {Herrero-Illana}, {Iwasawa}, {Komossa}, {Lena}, {Loiseau}, {Perez-Torres},
  {Piconcelli}, \& {Volonteri}}]{DeRosa2019}
{De Rosa}, A., {Vignali}, C., {Bogdanovi{\'c}}, T., {et~al.} 2019, \nar, 86,
  101525, \dodoi{10.1016/j.newar.2020.101525}

\bibitem[{{Deane} {et~al.}(2013){Deane}, {Rawlings}, {Garrett}, {Heywood},
  {Jarvis}, {Kl{\"o}ckner}, {Marshall}, \& {McKean}}]{Deane2013}
{Deane}, R.~P., {Rawlings}, S., {Garrett}, M.~A., {et~al.} 2013, \mnras, 434,
  3322, \dodoi{10.1093/mnras/stt1241}

\bibitem[{{Deane} {et~al.}(2014){Deane}, {Paragi}, {Jarvis}, {Coriat},
  {Bernardi}, {Fender}, {Frey}, {Heywood}, {Kl{\"o}ckner}, {Grainge}, \&
  {Rumsey}}]{Deane2014}
{Deane}, R.~P., {Paragi}, Z., {Jarvis}, M.~J., {et~al.} 2014, \nat, 511, 57,
  \dodoi{10.1038/nature13454}

\bibitem[{{Dickey} \& {Lockman}(1990)}]{Dickey1990}
{Dickey}, J.~M., \& {Lockman}, F.~J. 1990, \araa, 28, 215,
  \dodoi{10.1146/annurev.aa.28.090190.001243}

\bibitem[{{Dye} {et~al.}(2018){Dye}, {Furlanetto}, {Dunne}, {Eales},
  {Negrello}, {Nayyeri}, {van der Werf}, {Serjeant}, {Farrah},
  {Micha{\l}owski}, {Baes}, {Marchetti}, {Cooray}, {Riechers}, \&
  {Amvrosiadis}}]{Dye2018}
{Dye}, S., {Furlanetto}, C., {Dunne}, L., {et~al.} 2018, \mnras, 476, 4383,
  \dodoi{10.1093/mnras/sty513}

\bibitem[{{Enoki} {et~al.}(2004){Enoki}, {Inoue}, {Nagashima}, \&
  {Sugiyama}}]{Enoki2004}
{Enoki}, M., {Inoue}, K.~T., {Nagashima}, M., \& {Sugiyama}, N. 2004, \apj,
  615, 19, \dodoi{10.1086/424475}

\bibitem[{{Fabian}(2012)}]{Fabian2012}
{Fabian}, A.~C. 2012, \araa, 50, 455,
  \dodoi{10.1146/annurev-astro-081811-125521}

\bibitem[{{Fassnacht} \& {Cohen}(1998)}]{Fassnacht1998}
{Fassnacht}, C.~D., \& {Cohen}, J.~G. 1998, \aj, 115, 377,
  \dodoi{10.1086/300219}

\bibitem[{{Fassnacht} {et~al.}(2008){Fassnacht}, {Kocevski}, {Auger}, {Lubin},
  {Neureuther}, {Jeltema}, {Mulchaey}, \& {McKean}}]{Fassnacht2008}
{Fassnacht}, C.~D., {Kocevski}, D.~D., {Auger}, M.~W., {et~al.} 2008, \apj,
  681, 1017, \dodoi{10.1086/588087}

\bibitem[{{Fassnacht} \& {Lubin}(2002)}]{Fassnacht2002}
{Fassnacht}, C.~D., \& {Lubin}, L.~M. 2002, \aj, 123, 627,
  \dodoi{10.1086/338648}

\bibitem[{{Fassnacht} {et~al.}(1999){Fassnacht}, {Pearson}, {Readhead},
  {Browne}, {Koopmans}, {Myers}, \& {Wilkinson}}]{Fassnacht1999}
{Fassnacht}, C.~D., {Pearson}, T.~J., {Readhead}, A.~C.~S., {et~al.} 1999,
  \apj, 527, 498, \dodoi{10.1086/308118}

\bibitem[{{Fassnacht} {et~al.}(1996){Fassnacht}, {Womble}, {Neugebauer},
  {Browne}, {Readhead}, {Matthews}, \& {Pearson}}]{Fassnacht1996}
{Fassnacht}, C.~D., {Womble}, D.~S., {Neugebauer}, G., {et~al.} 1996, \apjl,
  460, L103, \dodoi{10.1086/309984}

\bibitem[{{Frey} {et~al.}(2015){Frey}, {Paragi}, {Fogasy}, \&
  {Gurvits}}]{Frey2015}
{Frey}, S., {Paragi}, Z., {Fogasy}, J.~O., \& {Gurvits}, L.~I. 2015, \mnras,
  446, 2921, \dodoi{10.1093/mnras/stu2294}

\bibitem[{{Fruscione} {et~al.}(2006){Fruscione}, {McDowell}, {Allen},
  {Brickhouse}, {Burke}, {Davis}, {Durham}, {Elvis}, {Galle}, {Harris},
  {Huenemoerder}, {Houck}, {Ishibashi}, {Karovska}, {Nicastro}, {Noble},
  {Nowak}, {Primini}, {Siemiginowska}, {Smith}, \& {Wise}}]{Fruscione2006}
{Fruscione}, A., {McDowell}, J.~C., {Allen}, G.~E., {et~al.} 2006, in Society
  of Photo-Optical Instrumentation Engineers (SPIE) Conference Series, Vol.
  6270, Society of Photo-Optical Instrumentation Engineers (SPIE) Conference
  Series, ed. D.~R. {Silva} \& R.~E. {Doxsey}, 62701V,
  \dodoi{10.1117/12.671760}

\bibitem[{{Fu} {et~al.}(2012){Fu}, {Yan}, {Myers}, {Stockton}, {Djorgovski},
  {Aldering}, \& {Rich}}]{Fu2012}
{Fu}, H., {Yan}, L., {Myers}, A.~D., {et~al.} 2012, \apj, 745, 67,
  \dodoi{10.1088/0004-637X/745/1/67}

\bibitem[{{Gilli} {et~al.}(2009){Gilli}, {Zamorani}, {Miyaji}, {Silverman},
  {Brusa}, {Mainieri}, {Cappelluti}, {Daddi}, {Porciani}, {Pozzetti}, {Civano},
  {Comastri}, {Finoguenov}, {Fiore}, {Salvato}, {Vignali}, {Hasinger}, {Lilly},
  {Impey}, {Trump}, {Capak}, {McCracken}, {Scoville}, {Taniguchi}, {Carollo},
  {Contini}, {Kneib}, {Le Fevre}, {Renzini}, {Scodeggio}, {Bardelli},
  {Bolzonella}, {Bongiorno}, {Caputi}, {Cimatti}, {Coppa}, {Cucciati}, {de La
  Torre}, {de Ravel}, {Franzetti}, {Garilli}, {Iovino}, {Kampczyk}, {Knobel},
  {Kova{\v{c}}}, {Lamareille}, {Le Borgne}, {Le Brun}, {Maier}, {Mignoli},
  {Pell{\`o}}, {Peng}, {Perez Montero}, {Ricciardelli}, {Tanaka}, {Tasca},
  {Tresse}, {Vergani}, {Zucca}, {Abbas}, {Bottini}, {Cappi}, {Cassata},
  {Fumana}, {Guzzo}, {Leauthaud}, {Maccagni}, {Marinoni}, {Memeo}, {Meneux},
  {Oesch}, {Scaramella}, \& {Walcher}}]{Gilli2009}
{Gilli}, R., {Zamorani}, G., {Miyaji}, T., {et~al.} 2009, \aap, 494, 33,
  \dodoi{10.1051/0004-6361:200810821}

\bibitem[{{G{\"u}ltekin} \& {Miller}(2012)}]{Gultekin2012}
{G{\"u}ltekin}, K., \& {Miller}, J.~M. 2012, \apj, 761, 90,
  \dodoi{10.1088/0004-637X/761/2/90}

\bibitem[{{Haardt} \& {Maraschi}(1991)}]{Haardt1991}
{Haardt}, F., \& {Maraschi}, L. 1991, \apjl, 380, L51, \dodoi{10.1086/186171}

\bibitem[{{Harris} {et~al.}(2003){Harris}, {Biretta}, {Junor}, {Perlman},
  {Sparks}, \& {Wilson}}]{Harris2003}
{Harris}, D.~E., {Biretta}, J.~A., {Junor}, W., {et~al.} 2003, \apjl, 586, L41,
  \dodoi{10.1086/374773}

\bibitem[{{Hartley} {et~al.}(2019){Hartley}, {Jackson}, {Sluse}, {Stacey}, \&
  {Vives-Arias}}]{Hartley2019}
{Hartley}, P., {Jackson}, N., {Sluse}, D., {Stacey}, H.~R., \& {Vives-Arias},
  H. 2019, \mnras, 485, 3009, \dodoi{10.1093/mnras/stz510}

\bibitem[{{Helmi} {et~al.}(2018){Helmi}, {Babusiaux}, {Koppelman}, {Massari},
  {Veljanoski}, \& {Brown}}]{Helmi2018}
{Helmi}, A., {Babusiaux}, C., {Koppelman}, H.~H., {et~al.} 2018, \nat, 563, 85,
  \dodoi{10.1038/s41586-018-0625-x}

\bibitem[{{Hopkins} {et~al.}(2006){Hopkins}, {Hernquist}, {Cox}, {Di Matteo},
  {Robertson}, \& {Springel}}]{Hopkins2006}
{Hopkins}, P.~F., {Hernquist}, L., {Cox}, T.~J., {et~al.} 2006, \apjs, 163, 1,
  \dodoi{10.1086/499298}

\bibitem[{{Hopkins} {et~al.}(2008){Hopkins}, {Hernquist}, {Cox}, \&
  {Kere{\v{s}}}}]{Hopkins2008}
{Hopkins}, P.~F., {Hernquist}, L., {Cox}, T.~J., \& {Kere{\v{s}}}, D. 2008,
  \apjs, 175, 356, \dodoi{10.1086/524362}

\bibitem[{{Hsueh} {et~al.}(2017){Hsueh}, {Oldham}, {Spingola}, {Vegetti},
  {Fassnacht}, {Auger}, {Koopmans}, {McKean}, \& {Lagattuta}}]{Hsueh2017}
{Hsueh}, J.~W., {Oldham}, L., {Spingola}, C., {et~al.} 2017, \mnras, 469, 3713,
  \dodoi{10.1093/mnras/stx1082}

\bibitem[{{Ighina} {et~al.}(2021){Ighina}, {Moretti}, {Tavecchio},
  {Caccianiga}, {Belladitta}, {Dallacasa}, {Della Ceca}, {Sbarrato}, \&
  {Spingola}}]{Ighina2021}
{Ighina}, L., {Moretti}, A., {Tavecchio}, F., {et~al.} 2021, arXiv e-prints,
  arXiv:2111.08632.
\newblock \doarXiv{2111.08632}

\bibitem[{{Jerius} {et~al.}(2004){Jerius}, {Cohen}, {Edgar}, {Freeman},
  {Gaetz}, {Hughes}, {Nguyen}, {Podgorski}, {Tibbetts}, {Van Speybroeck}, \&
  {Zhao}}]{Jerius2004}
{Jerius}, D.~H., {Cohen}, L., {Edgar}, R.~J., {et~al.} 2004, in Society of
  Photo-Optical Instrumentation Engineers (SPIE) Conference Series, Vol. 5165,
  X-Ray and Gamma-Ray Instrumentation for Astronomy XIII, ed. K.~A. {Flanagan}
  \& O.~H.~W. {Siegmund}, 402--410, \dodoi{10.1117/12.509378}

\bibitem[{{Joye} \& {Mandel}(2003)}]{Joye2003}
{Joye}, W.~A., \& {Mandel}, E. 2003, in Astronomical Society of the Pacific
  Conference Series, Vol. 295, Astronomical Data Analysis Software and Systems
  XII, ed. H.~E. {Payne}, R.~I. {Jedrzejewski}, \& R.~N. {Hook}, 489

\bibitem[{{Keeton}(2001{\natexlab{a}})}]{Keeton2001a}
{Keeton}, C.~R. 2001{\natexlab{a}}, arXiv e-prints, astro.
\newblock \doarXiv{astro-ph/0102340}

\bibitem[{{Keeton}(2001{\natexlab{b}})}]{Keeton2001b}
---. 2001{\natexlab{b}}, arXiv e-prints, astro.
\newblock \doarXiv{astro-ph/0102341}

\bibitem[{{Kim} {et~al.}(2016){Kim}, {Evans}, {Stierwalt}, \&
  {Privon}}]{Kim2016}
{Kim}, D.~C., {Evans}, A.~S., {Stierwalt}, S., \& {Privon}, G.~C. 2016, \apj,
  824, 122, \dodoi{10.3847/0004-637X/824/2/122}

\bibitem[{{Kong} {et~al.}(2010){Kong}, {Heinke}, {di Stefano}, {Cohn},
  {Lugger}, {Barmby}, {Lewin}, \& {Primini}}]{Kong2010}
{Kong}, A.~K.~H., {Heinke}, C.~O., {di Stefano}, R., {et~al.} 2010, \mnras,
  407, L84, \dodoi{10.1111/j.1745-3933.2010.00910.x}

\bibitem[{{Koopmans} {et~al.}(2006){Koopmans}, {Treu}, {Bolton}, {Burles}, \&
  {Moustakas}}]{Koopmans2006}
{Koopmans}, L. V.~E., {Treu}, T., {Bolton}, A.~S., {Burles}, S., \&
  {Moustakas}, L.~A. 2006, \apj, 649, 599, \dodoi{10.1086/505696}

\bibitem[{{Koopmans} {et~al.}(2003){Koopmans}, {Treu}, {Fassnacht},
  {Blandford}, \& {Surpi}}]{Koopmans2003}
{Koopmans}, L.~V.~E., {Treu}, T., {Fassnacht}, C.~D., {Blandford}, R.~D., \&
  {Surpi}, G. 2003, \apj, 599, 70, \dodoi{10.1086/379226}

\bibitem[{{Koss} {et~al.}(2012){Koss}, {Mushotzky}, {Treister}, {Veilleux},
  {Vasudevan}, \& {Trippe}}]{Koss2012}
{Koss}, M., {Mushotzky}, R., {Treister}, E., {et~al.} 2012, \apjl, 746, L22,
  \dodoi{10.1088/2041-8205/746/2/L22}

\bibitem[{{Lemon} {et~al.}(2021){Lemon}, {Millon}, {Sluse}, {Courbin}, {Auger},
  {Chan}, {Paic}, \& {Agnello}}]{Lemon2021}
{Lemon}, C., {Millon}, M., {Sluse}, D., {et~al.} 2021, arXiv e-prints,
  arXiv:2109.01144.
\newblock \doarXiv{2109.01144}

\bibitem[{{Lena} {et~al.}(2014){Lena}, {Robinson}, {Marconi}, {Axon},
  {Capetti}, {Merritt}, \& {Batcheldor}}]{Lena2014}
{Lena}, D., {Robinson}, A., {Marconi}, A., {et~al.} 2014, \apj, 795, 146,
  \dodoi{10.1088/0004-637X/795/2/146}

\bibitem[{{Libralato} {et~al.}(2021){Libralato}, {Lennon}, {Bellini}, {van der
  Marel}, {Clark}, {Najarro}, {Patrick}, {Anderson}, {Bedin}, {Crowther}, {de
  Mink}, {Evans}, {Platais}, {Sabbi}, \& {Sohn}}]{Libralato2021}
{Libralato}, M., {Lennon}, D.~J., {Bellini}, A., {et~al.} 2021, \mnras, 500,
  3213, \dodoi{10.1093/mnras/staa3329}

\bibitem[{{Lindegren}(2020)}]{Lindegren2020}
{Lindegren}, L. 2020, \aap, 633, A1, \dodoi{10.1051/0004-6361/201936161}

\bibitem[{{Liu} {et~al.}(2011){Liu}, {Shen}, {Strauss}, \& {Hao}}]{Liu2011}
{Liu}, X., {Shen}, Y., {Strauss}, M.~A., \& {Hao}, L. 2011, \apj, 737, 101,
  \dodoi{10.1088/0004-637X/737/2/101}

\bibitem[{{Lodato} {et~al.}(2009){Lodato}, {Nayakshin}, {King}, \&
  {Pringle}}]{Lodato2009}
{Lodato}, G., {Nayakshin}, S., {King}, A.~R., \& {Pringle}, J.~E. 2009, \mnras,
  398, 1392, \dodoi{10.1111/j.1365-2966.2009.15179.x}

\bibitem[{{Lopez Armengol} {et~al.}(2021){Lopez Armengol}, {Combi},
  {Campanelli}, {Noble}, {Krolik}, {Bowen}, {Avara}, {Mewes}, \&
  {Nakano}}]{LopezArmengol2021}
{Lopez Armengol}, F.~G., {Combi}, L., {Campanelli}, M., {et~al.} 2021, \apj,
  913, 16, \dodoi{10.3847/1538-4357/abf0af}

\bibitem[{{Madau} \& {Quataert}(2004)}]{Madau2004}
{Madau}, P., \& {Quataert}, E. 2004, \apjl, 606, L17, \dodoi{10.1086/421017}

\bibitem[{{Mangat} {et~al.}(2021){Mangat}, {McKean}, {Brilenkov}, {Hartley},
  {Stacey}, {Vegetti}, \& {Wen}}]{Mangat2021}
{Mangat}, C.~S., {McKean}, J.~P., {Brilenkov}, R., {et~al.} 2021, arXiv
  e-prints, arXiv:2109.03253.
\newblock \doarXiv{2109.03253}

\bibitem[{{Massardi} {et~al.}(2018){Massardi}, {Enia}, {Negrello}, {Mancuso},
  {Lapi}, {Vignali}, {Gilli}, {Burkutean}, {Danese}, \& {Zotti}}]{Massardi2018}
{Massardi}, M., {Enia}, A.~F.~M., {Negrello}, M., {et~al.} 2018, \aap, 610,
  A53, \dodoi{10.1051/0004-6361/201731751}

\bibitem[{{Massari} {et~al.}(2018){Massari}, {Breddels}, {Helmi}, {Posti},
  {Brown}, \& {Tolstoy}}]{Massari2018}
{Massari}, D., {Breddels}, M.~A., {Helmi}, A., {et~al.} 2018, Nature Astronomy,
  2, 156, \dodoi{10.1038/s41550-017-0322-y}

\bibitem[{{Massari} {et~al.}(2020){Massari}, {Helmi}, {Mucciarelli}, {Sales},
  {Spina}, \& {Tolstoy}}]{Massari2020}
{Massari}, D., {Helmi}, A., {Mucciarelli}, A., {et~al.} 2020, \aap, 633, A36,
  \dodoi{10.1051/0004-6361/201935613}

\bibitem[{{McKean} {et~al.}(2015){McKean}, {Jackson}, {Vegetti}, {Rybak},
  {Serjeant}, {Koopmans}, {Metcalf}, {Fassnacht}, {Marshall}, \&
  {Pandey-Pommier}}]{McKean2015}
{McKean}, J., {Jackson}, N., {Vegetti}, S., {et~al.} 2015, in Advancing
  Astrophysics with the Square Kilometre Array (AASKA14), 84.
\newblock \doarXiv{1502.03362}

\bibitem[{{Momcheva} {et~al.}(2015){Momcheva}, {Williams}, {Cool}, {Keeton}, \&
  {Zabludoff}}]{Momcheva2015}
{Momcheva}, I.~G., {Williams}, K.~A., {Cool}, R.~J., {Keeton}, C.~R., \&
  {Zabludoff}, A.~I. 2015, \apjs, 219, 29, \dodoi{10.1088/0067-0049/219/2/29}

\bibitem[{{Momjian} {et~al.}(2021){Momjian}, {Ba{\~n}ados}, {Carilli},
  {Walter}, \& {Mazzucchelli}}]{Momjian2021}
{Momjian}, E., {Ba{\~n}ados}, E., {Carilli}, C.~L., {Walter}, F., \&
  {Mazzucchelli}, C. 2021, \aj, 161, 207, \dodoi{10.3847/1538-3881/abe6ae}

\bibitem[{{M{\"u}ller-S{\'a}nchez} {et~al.}(2015){M{\"u}ller-S{\'a}nchez},
  {Comerford}, {Nevin}, {Barrows}, {Cooper}, \& {Greene}}]{Muller-Sanchez2015}
{M{\"u}ller-S{\'a}nchez}, F., {Comerford}, J.~M., {Nevin}, R., {et~al.} 2015,
  \apj, 813, 103, \dodoi{10.1088/0004-637X/813/2/103}

\bibitem[{{Nair} {et~al.}(2005){Nair}, {Jin}, \& {Garrett}}]{Nair2005}
{Nair}, S., {Jin}, C., \& {Garrett}, M.~A. 2005, \mnras, 362, 1157,
  \dodoi{10.1111/j.1365-2966.2005.09355.x}

\bibitem[{{Orosz} \& {Frey}(2013)}]{Orosz2013}
{Orosz}, G., \& {Frey}, S. 2013, \aap, 553, A13,
  \dodoi{10.1051/0004-6361/201321279}

\bibitem[{{Padovani} {et~al.}(2017){Padovani}, {Alexander}, {Assef}, {De
  Marco}, {Giommi}, {Hickox}, {Richards}, {Smol{\v{c}}i{\'c}},
  {Hatziminaoglou}, {Mainieri}, \& {Salvato}}]{Padovani2017}
{Padovani}, P., {Alexander}, D.~M., {Assef}, R.~J., {et~al.} 2017, \aapr, 25,
  2, \dodoi{10.1007/s00159-017-0102-9}

\bibitem[{{Perger} {et~al.}(2021){Perger}, {Frey}, {Schwartz}, {Gab{\'a}nyi},
  {Gurvits}, \& {Paragi}}]{Perger2021}
{Perger}, K., {Frey}, S., {Schwartz}, D.~A., {et~al.} 2021, arXiv e-prints,
  arXiv:2105.06307.
\newblock \doarXiv{2105.06307}

\bibitem[{{Perger} {et~al.}(2018){Perger}, {Frey}, {Gab{\'a}nyi}, {An},
  {Britzen}, {Cao}, {Cseh}, {Dennett-Thorpe}, {Gurvits}, {Hong}, {Hook},
  {Paragi}, {Schilizzi}, {Yang}, \& {Zhang}}]{Perger2018}
{Perger}, K., {Frey}, S., {Gab{\'a}nyi}, K.~{\'E}., {et~al.} 2018, \mnras, 477,
  1065, \dodoi{10.1093/mnras/sty837}

\bibitem[{{Planck Collaboration} {et~al.}(2016){Planck Collaboration}, {Ade},
  {Aghanim}, {Arnaud}, {Ashdown}, {Aumont}, {Baccigalupi}, {Banday},
  {Barreiro}, {Bartlett}, {Bartolo}, {Battaner}, {Battye}, {Benabed},
  {Beno{\^\i}t}, {Benoit-L{\'e}vy}, {Bernard}, {Bersanelli}, {Bielewicz},
  {Bock}, {Bonaldi}, {Bonavera}, {Bond}, {Borrill}, {Bouchet}, {Boulanger},
  {Bucher}, {Burigana}, {Butler}, {Calabrese}, {Cardoso}, {Catalano},
  {Challinor}, {Chamballu}, {Chary}, {Chiang}, {Chluba}, {Christensen},
  {Church}, {Clements}, {Colombi}, {Colombo}, {Combet}, {Coulais}, {Crill},
  {Curto}, {Cuttaia}, {Danese}, {Davies}, {Davis}, {de Bernardis}, {de Rosa},
  {de Zotti}, {Delabrouille}, {D{\'e}sert}, {Di Valentino}, {Dickinson},
  {Diego}, {Dolag}, {Dole}, {Donzelli}, {Dor{\'e}}, {Douspis}, {Ducout},
  {Dunkley}, {Dupac}, {Efstathiou}, {Elsner}, {En{\ss}lin}, {Eriksen},
  {Farhang}, {Fergusson}, {Finelli}, {Forni}, {Frailis}, {Fraisse},
  {Franceschi}, {Frejsel}, {Galeotta}, {Galli}, {Ganga}, {Gauthier}, {Gerbino},
  {Ghosh}, {Giard}, {Giraud-H{\'e}raud}, {Giusarma}, {Gjerl{\o}w},
  {Gonz{\'a}lez-Nuevo}, {G{\'o}rski}, {Gratton}, {Gregorio}, {Gruppuso},
  {Gudmundsson}, {Hamann}, {Hansen}, {Hanson}, {Harrison}, {Helou},
  {Henrot-Versill{\'e}}, {Hern{\'a}ndez-Monteagudo}, {Herranz}, {Hildebrandt},
  {Hivon}, {Hobson}, {Holmes}, {Hornstrup}, {Hovest}, {Huang}, {Huffenberger},
  {Hurier}, {Jaffe}, {Jaffe}, {Jones}, {Juvela}, {Keih{\"a}nen}, {Keskitalo},
  {Kisner}, {Kneissl}, {Knoche}, {Knox}, {Kunz}, {Kurki-Suonio}, {Lagache},
  {L{\"a}hteenm{\"a}ki}, {Lamarre}, {Lasenby}, {Lattanzi}, {Lawrence}, {Leahy},
  {Leonardi}, {Lesgourgues}, {Levrier}, {Lewis}, {Liguori}, {Lilje},
  {Linden-V{\o}rnle}, {L{\'o}pez-Caniego}, {Lubin}, {Mac{\'\i}as-P{\'e}rez},
  {Maggio}, {Maino}, {Mandolesi}, {Mangilli}, {Marchini}, {Maris}, {Martin},
  {Martinelli}, {Mart{\'\i}nez-Gonz{\'a}lez}, {Masi}, {Matarrese}, {McGehee},
  {Meinhold}, {Melchiorri}, {Melin}, {Mendes}, {Mennella}, {Migliaccio},
  {Millea}, {Mitra}, {Miville-Desch{\^e}nes}, {Moneti}, {Montier}, {Morgante},
  {Mortlock}, {Moss}, {Munshi}, {Murphy}, {Naselsky}, {Nati}, {Natoli},
  {Netterfield}, {N{\o}rgaard-Nielsen}, {Noviello}, {Novikov}, {Novikov},
  {Oxborrow}, {Paci}, {Pagano}, {Pajot}, {Paladini}, {Paoletti}, {Partridge},
  {Pasian}, {Patanchon}, {Pearson}, {Perdereau}, {Perotto}, {Perrotta},
  {Pettorino}, {Piacentini}, {Piat}, {Pierpaoli}, {Pietrobon}, {Plaszczynski},
  {Pointecouteau}, {Polenta}, {Popa}, {Pratt}, {Pr{\'e}zeau}, {Prunet},
  {Puget}, {Rachen}, {Reach}, {Rebolo}, {Reinecke}, {Remazeilles}, {Renault},
  {Renzi}, {Ristorcelli}, {Rocha}, {Rosset}, {Rossetti}, {Roudier},
  {Rouill{\'e} d'Orfeuil}, {Rowan-Robinson}, {Rubi{\~n}o-Mart{\'\i}n},
  {Rusholme}, {Said}, {Salvatelli}, {Salvati}, {Sandri}, {Santos},
  {Savelainen}, {Savini}, {Scott}, {Seiffert}, {Serra}, {Shellard}, {Spencer},
  {Spinelli}, {Stolyarov}, {Stompor}, {Sudiwala}, {Sunyaev}, {Sutton},
  {Suur-Uski}, {Sygnet}, {Tauber}, {Terenzi}, {Toffolatti}, {Tomasi},
  {Tristram}, {Trombetti}, {Tucci}, {Tuovinen}, {T{\"u}rler}, {Umana},
  {Valenziano}, {Valiviita}, {Van Tent}, {Vielva}, {Villa}, {Wade}, {Wandelt},
  {Wehus}, {White}, {White}, {Wilkinson}, {Yvon}, {Zacchei}, \&
  {Zonca}}]{Planck2016}
{Planck Collaboration}, {Ade}, P.~A.~R., {Aghanim}, N., {et~al.} 2016, \aap,
  594, A13, \dodoi{10.1051/0004-6361/201525830}

\bibitem[{{Ponti} {et~al.}(2019){Ponti}, {Hofmann}, {Churazov}, {Morris},
  {Haberl}, {Nandra}, {Terrier}, {Clavel}, \& {Goldwurm}}]{Ponti2019}
{Ponti}, G., {Hofmann}, F., {Churazov}, E., {et~al.} 2019, \nat, 567, 347,
  \dodoi{10.1038/s41586-019-1009-6}

\bibitem[{{Rafikov}(2016)}]{Rafikov2016}
{Rafikov}, R.~R. 2016, \apj, 827, 111, \dodoi{10.3847/0004-637X/827/2/111}

\bibitem[{{Ratti} {et~al.}(2010){Ratti}, {Bassa}, {Torres}, {Kuiper},
  {Miller-Jones}, \& {Jonker}}]{Ratti2010}
{Ratti}, E.~M., {Bassa}, C.~G., {Torres}, M.~A.~P., {et~al.} 2010, \mnras, 408,
  1866, \dodoi{10.1111/j.1365-2966.2010.17252.x}

\bibitem[{{Reines} {et~al.}(2020){Reines}, {Condon}, {Darling}, \&
  {Greene}}]{Reines2020}
{Reines}, A.~E., {Condon}, J.~J., {Darling}, J., \& {Greene}, J.~E. 2020, \apj,
  888, 36, \dodoi{10.3847/1538-4357/ab4999}

\bibitem[{{Robitaille} \& {Bressert}(2012)}]{Robitaille2012}
{Robitaille}, T., \& {Bressert}, E. 2012, {APLpy: Astronomical Plotting Library
  in Python}.
\newblock \doeprint{1208.017}

\bibitem[{{Rosas-Guevara} {et~al.}(2019){Rosas-Guevara}, {Bower}, {McAlpine},
  {Bonoli}, \& {Tissera}}]{Rosas-guevara2019}
{Rosas-Guevara}, Y.~M., {Bower}, R.~G., {McAlpine}, S., {Bonoli}, S., \&
  {Tissera}, P.~B. 2019, \mnras, 483, 2712, \dodoi{10.1093/mnras/sty3251}

\bibitem[{{Rybak} {et~al.}(2020){Rybak}, {Hodge}, {Vegetti}, {van der Werf},
  {Andreani}, {Graziani}, \& {McKean}}]{Rybak2020}
{Rybak}, M., {Hodge}, J.~A., {Vegetti}, S., {et~al.} 2020, \mnras, 494, 5542,
  \dodoi{10.1093/mnras/staa879}

\bibitem[{{Schwartz} {et~al.}(2021){Schwartz}, {Spingola}, \&
  {Barnacka}}]{Schwartz2021}
{Schwartz}, D., {Spingola}, C., \& {Barnacka}, A. 2021, \apj, 917, 26,
  \dodoi{10.3847/1538-4357/ac0909}

\bibitem[{{Schwartz} {et~al.}(2000){Schwartz}, {Marshall}, {Lovell}, {Piner},
  {Tingay}, {Birkinshaw}, {Chartas}, {Elvis}, {Feigelson}, {Ghosh}, {Harris},
  {Hirabayashi}, {Hooper}, {Jauncey}, {Lanzetta}, {Mathur}, {Preston},
  {Tucker}, {Virani}, {Wilkes}, \& {Worrall}}]{Schwartz2000}
{Schwartz}, D.~A., {Marshall}, H.~L., {Lovell}, J.~E.~J., {et~al.} 2000, \apjl,
  540, 69, \dodoi{10.1086/312875}

\bibitem[{{Schwartz} {et~al.}(2020){Schwartz}, {Siemiginowska}, {Snios},
  {Worrall}, {Birkinshaw}, {Cheung}, {Marshall}, {Migliori}, {Wardle}, \&
  {Gobeille}}]{Schwartz2020}
{Schwartz}, D.~A., {Siemiginowska}, A., {Snios}, B., {et~al.} 2020, \apj, 904,
  57, \dodoi{10.3847/1538-4357/abbd99}

\bibitem[{{Shen} {et~al.}(2019){Shen}, {Hwang}, {Zakamska}, \&
  {Liu}}]{Shen2019}
{Shen}, Y., {Hwang}, H.-C., {Zakamska}, N., \& {Liu}, X. 2019, \apjl, 885, L4,
  \dodoi{10.3847/2041-8213/ab4b54}

\bibitem[{{Silva} {et~al.}(2021){Silva}, {Marchesini}, {Silverman}, {Martis},
  {Iono}, {Espada}, \& {Skelton}}]{Silva2021}
{Silva}, A., {Marchesini}, D., {Silverman}, J.~D., {et~al.} 2021, \apj, 909,
  124, \dodoi{10.3847/1538-4357/abdbb1}

\bibitem[{{Skipper} \& {Browne}(2018)}]{Skipper2018}
{Skipper}, C.~J., \& {Browne}, I. W.~A. 2018, \mnras, 475, 5179,
  \dodoi{10.1093/mnras/sty114}

\bibitem[{{Sluse} {et~al.}(2012){Sluse}, {Chantry}, {Magain}, {Courbin}, \&
  {Meylan}}]{Sluse2012}
{Sluse}, D., {Chantry}, V., {Magain}, P., {Courbin}, F., \& {Meylan}, G. 2012,
  \aap, 538, A99, \dodoi{10.1051/0004-6361/201015844}

\bibitem[{{Snios} {et~al.}(2021){Snios}, {Schwartz}, {Siemiginowska},
  {Sobolewska}, {Birkinshaw}, {Cheung}, {Gobeille}, {Marshall}, {Migliori},
  {Wardle}, \& {Worrall}}]{Snios2021}
{Snios}, B., {Schwartz}, D.~A., {Siemiginowska}, A., {et~al.} 2021, \apj, 914,
  130, \dodoi{10.3847/1538-4357/abfe64}

\bibitem[{{Somerville} \& {Dav{\'e}}(2015)}]{Somerville2015}
{Somerville}, R.~S., \& {Dav{\'e}}, R. 2015, \araa, 53, 51,
  \dodoi{10.1146/annurev-astro-082812-140951}

\bibitem[{{Spilker} {et~al.}(2015){Spilker}, {Aravena}, {Marrone},
  {B{\'e}thermin}, {Bothwell}, {Carlstrom}, {Chapman}, {Collier}, {de Breuck},
  {Fassnacht}, {Galvin}, {Gonzalez}, {Gonz{\'a}lez-L{\'o}pez}, {Grieve},
  {Hezaveh}, {Ma}, {Malkan}, {O'Brien}, {Rotermund}, {Strandet}, {Vieira},
  {Weiss}, \& {Wong}}]{Spilker2015}
{Spilker}, J.~S., {Aravena}, M., {Marrone}, D.~P., {et~al.} 2015, \apj, 811,
  124, \dodoi{10.1088/0004-637X/811/2/124}

\bibitem[{{Spingola} \& {Barnacka}(2020)}]{Spingola&Barnacka2020}
{Spingola}, C., \& {Barnacka}, A. 2020, \mnras, 494, 2312,
  \dodoi{10.1093/mnras/staa870}

\bibitem[{{Spingola} {et~al.}(2020{\natexlab{a}}){Spingola}, {Dallacasa},
  {Belladitta}, {Caccianiga}, {Giroletti}, {Moretti}, \&
  {Orienti}}]{Spingola2020}
{Spingola}, C., {Dallacasa}, D., {Belladitta}, S., {et~al.} 2020{\natexlab{a}},
  \aap, 643, L12, \dodoi{10.1051/0004-6361/202039458}

\bibitem[{{Spingola} {et~al.}(2018){Spingola}, {McKean}, {Auger}, {Fassnacht},
  {Koopmans}, {Lagattuta}, \& {Vegetti}}]{Spingola2018}
{Spingola}, C., {McKean}, J.~P., {Auger}, M.~W., {et~al.} 2018, \mnras, 478,
  4816, \dodoi{10.1093/mnras/sty1326}

\bibitem[{{Spingola} {et~al.}(2019){Spingola}, {McKean}, {Massari}, \&
  {Koopmans}}]{Spingola2019}
{Spingola}, C., {McKean}, J.~P., {Massari}, D., \& {Koopmans}, L.~V.~E. 2019,
  \aap, 630, A108, \dodoi{10.1051/0004-6361/201935427}

\bibitem[{{Spingola} {et~al.}(2020{\natexlab{b}}){Spingola}, {McKean},
  {Vegetti}, {Powell}, {Auger}, {Koopmans}, {Fassnacht}, {Lagattuta}, {Rizzo},
  {Stacey}, \& {Sweijen}}]{Spingola2020_gas}
{Spingola}, C., {McKean}, J.~P., {Vegetti}, S., {et~al.} 2020{\natexlab{b}},
  \mnras, 495, 2387, \dodoi{10.1093/mnras/staa1342}

\bibitem[{{Suyu} {et~al.}(2010){Suyu}, {Marshall}, {Auger}, {Hilbert},
  {Blandford}, {Koopmans}, {Fassnacht}, \& {Treu}}]{Suyu2010}
{Suyu}, S.~H., {Marshall}, P.~J., {Auger}, M.~W., {et~al.} 2010, \apj, 711,
  201, \dodoi{10.1088/0004-637X/711/1/201}

\bibitem[{{Tomsick} {et~al.}(2020){Tomsick}, {Bodaghee}, {Chaty}, {Clavel},
  {Fornasini}, {Hare}, {Krivonos}, {Rahoui}, \& {Rodriguez}}]{Tomsick2020}
{Tomsick}, J.~A., {Bodaghee}, A., {Chaty}, S., {et~al.} 2020, \apj, 889, 53,
  \dodoi{10.3847/1538-4357/ab5fd2}

\bibitem[{{Tomsick} {et~al.}(2021){Tomsick}, {Coughenour}, {Hare}, {Krivonos},
  {Bodaghee}, {Chaty}, {Clavel}, {Fornasini}, {Rodriguez}, \&
  {Shaw}}]{Tomsick2021}
{Tomsick}, J.~A., {Coughenour}, B.~M., {Hare}, J., {et~al.} 2021, \apj, 914,
  48, \dodoi{10.3847/1538-4357/abfa1a}

\bibitem[{{Tremblay} {et~al.}(2016){Tremblay}, {Taylor}, {Ortiz}, {Tremblay},
  {Helmboldt}, \& {Romani}}]{Tremblay2016}
{Tremblay}, S.~E., {Taylor}, G.~B., {Ortiz}, A.~A., {et~al.} 2016, \mnras, 459,
  820, \dodoi{10.1093/mnras/stw592}

\bibitem[{{Vito} {et~al.}(2019){Vito}, {Brandt}, {Bauer}, {Calura}, {Gilli},
  {Luo}, {Shemmer}, {Vignali}, {Zamorani}, {Brusa}, {Civano}, {Comastri}, \&
  {Nanni}}]{Vito2019}
{Vito}, F., {Brandt}, W.~N., {Bauer}, F.~E., {et~al.} 2019, \aap, 630, A118,
  \dodoi{10.1051/0004-6361/201936217}

\bibitem[{{Vito} {et~al.}(2021){Vito}, {Brandt}, {Ricci}, {Congiu}, {Connor},
  {Ba{\~n}ados}, {Bauer}, {Gilli}, {Luo}, {Mazzucchelli}, {Mignoli}, {Shemmer},
  {Vignali}, {Calura}, {Comastri}, {Decarli}, {Gallerani}, {Nanni}, {Brusa},
  {Cappelluti}, {Civano}, \& {Zamorani}}]{Vito2021}
{Vito}, F., {Brandt}, W.~N., {Ricci}, F., {et~al.} 2021, \aap, 649, A133,
  \dodoi{10.1051/0004-6361/202140399}

\bibitem[{{Volonteri} \& {Madau}(2008)}]{Volonteri2008}
{Volonteri}, M., \& {Madau}, P. 2008, \apjl, 687, L57, \dodoi{10.1086/593353}

\bibitem[{{Wang} {et~al.}(2009){Wang}, {Chen}, {Hu}, {Mao}, {Zhang}, \&
  {Bian}}]{Wang2009}
{Wang}, J.-M., {Chen}, Y.-M., {Hu}, C., {et~al.} 2009, \apjl, 705, L76,
  \dodoi{10.1088/0004-637X/705/1/L76}

\bibitem[{{Wang} {et~al.}(2016){Wang}, {Liu}, {Qiu}, {Bai}, {Yang}, {Guo}, \&
  {Zhang}}]{Wang2016}
{Wang}, S., {Liu}, J., {Qiu}, Y., {et~al.} 2016, VizieR Online Data Catalog,
  J/ApJS/224/40

\bibitem[{{Wilks}(1938)}]{Wilks1938}
{Wilks}, S.~S. 1938, Ann. Math Stat., 9, 60

\bibitem[{{Wilson} {et~al.}(2017){Wilson}, {Zabludoff}, {Keeton}, {Wong},
  {Williams}, {French}, \& {Momcheva}}]{Wilson2017}
{Wilson}, M.~L., {Zabludoff}, A.~I., {Keeton}, C.~R., {et~al.} 2017, \apj, 850,
  94, \dodoi{10.3847/1538-4357/aa9653}

\bibitem[{{Worrall}(2009)}]{Worrall2009}
{Worrall}, D.~M. 2009, \aapr, 17, 1, \dodoi{10.1007/s00159-008-0016-7}

\bibitem[{{Yan} {et~al.}(2015){Yan}, {Lu}, {Dai}, \& {Yu}}]{Yan2015}
{Yan}, C.-S., {Lu}, Y., {Dai}, X., \& {Yu}, Q. 2015, \apj, 809, 117,
  \dodoi{10.1088/0004-637X/809/2/117}

\bibitem[{{Zhang} {et~al.}(2021){Zhang}, {An}, {Frey}, {Yang}, {Krezinger},
  {Titov}, {Melnikov}, {de Vicente}, {Shu}, \& {Wang}}]{Zhang2021}
{Zhang}, Y., {An}, T., {Frey}, S., {et~al.} 2021, \mnras, 507, 3736,
  \dodoi{10.1093/mnras/stab2289}

\end{thebibliography}
\bibliographystyle{aasjournal}

\end{document}